\documentclass{aa}  
\usepackage{graphicx}
\usepackage{txfonts}
\begin{document}
\title{The substellar mass function in $\sigma$~Orionis. II}
\subtitle{Optical, near-infrared and IRAC/{\em Spitzer} photometry of young
cluster brown dwarfs and planetary-mass objects} 
\author{J. A. Caballero\inst{1,2}
     \and
     V. J. S. B\'ejar\inst{3}
     \and
     R. Rebolo\inst{1,4}
     \and
     J. Eisl\"offel\inst{5}
     \and
     M. R. Zapatero Osorio\inst{1,6}     
     \and
     R. Mundt\inst{2}
     \and
     D. Barrado y Navascu\'es\inst{6}
     \and
     G. Bihain\inst{1,4}
     \and
     C. A. L. Bailer-Jones\inst{2}
     \and
     T. Forveille\inst{7,8}
     \and
     E. L. Mart\'{\i}n\inst{1,9}
}
\offprints{Jos\'e Antonio Caballero (Alexander von Humboldt Fellow at the
Max-Planck-Institut f\"ur Astronomie), \email{caballero@mpia.de}}
\institute{Instituto de Astrof\'{\i}sica de Canarias, Avenida V\'{\i}a L\'actea,
E-38200 La Laguna, Tenerife, Islas Canarias, Spain 
\and 
Max-Planck-Institut f\"ur Astronomie, K\"onigstuhl 17, D-69117 Heidelberg,
Germany 
\and 
Proyecto Gran Telescopio Canarias, Instituto de Astrof\'{\i}sica de Canarias 
\and 
Consejo Superior de Investigaciones Cient\'{\i}ficas, Spain
\and 
Th\"uringer Landessternwarte, Sternwarte 5, D-07778 Tautenburg, Germany
\and 
LAEFF-INTA, P.O. Box 50727, E-28080, Madrid, Spain
\and 
Canada-France-Hawaii Telescope Corporation, 65-1238 Mamalahoa Highway, Kamuela,
HI96743, Hawai'i, USA  
\and
Laboratoire d'Astrophysique, Observatoire de Grenoble, BP 53, 38041 Grenoble
Cedex 9, France 
\and 
University of Central Florida, Dept. of Physics, P.O. Box 162385, Orlando, FL
32816-2385, USA  
}
\date{Received 21 December 2006; accepted 12 April 2007}

% \abstract{}{}{}{}{} 
% 5 {} token are mandatory
\abstract
{}   
{We investigate the mass function in the substellar domain down to a
few Jupiter masses in the young $\sigma$~Orionis open cluster (3$\pm$2\,Ma,
$d$ = 360$^{+70}_{-60}$\,pc).} 
{We have performed a deep $IJ$-band search, covering an area of 
790\,arcmin$^2$ close to the cluster centre. 
This survey was complemented with an infrared follow-up in the $HK_{\rm s}$- and 
Spitzer 3.6--8.0\,$\mu$m-bands.
Using colour-magnitude diagrams, we have selected 49 candidate cluster
members in the magnitude interval 16.1\,mag $< I <$ 23.0\,mag.}    
{Accounting for flux excesses at 8.0\,$\mu$m and previously known spectral
features of youth, 30 objects are bona fide cluster members. 
Four are first identified from our optical-near infrared data.
Eleven have most probable masses below the deuterium burning limit and are
classified as planetary-mass object candidates.
The slope of the substellar mass spectrum ($\Delta N / \Delta {\rm M} \approx a
{\rm M}^{-\alpha}$) in the mass interval 0.11\,$M_\odot$
$< {\rm M} <$ 0.006\,$M_\odot$ is $\alpha = +0.6 \pm 0.2$. 
Any opacity mass-limit, if these objects form via fragmentation, may lie below
0.006\,$M_\odot$. 
The frequency of $\sigma$~Orionis brown dwarfs with circumsubstellar discs is
47$\pm$15\,\%.} 
{The continuity in the mass function and in the frequency of discs suggests that
very low-mass stars and substellar objects, even below the deuterium-burning
mass limit, may share the same formation mechanism.} 
\maketitle
%
%________________________________________________________________

\section{Introduction}
\label{intro}

The increasing sensitivity of photometric searches in young open clusters and
star-forming regions {(1--10\,Ma)} has led to the direct detection of substellar
objects with theoretical masses below the deuterium burning mass limit
(0.013\,$M_\odot$ for solar metallicity; Chabrier \& Baraffe 2000).   
This mass limit has been used by several authors as a boundary to separate brown
dwarfs from planetary-mass objects (PMOs).  
The first directly imaged PMOs were found as isolated objects in very young open
clusters (Zapatero Osorio et al. 2000; Lucas \& Roche 2000; Najita, Tiede \&
Carr 2000).
They are sometimes called isolated planetary-mass objects (IPMOs) to
differentiate them from recently discovered PMOs orbiting stars and brown dwarfs
(Chauvin et al. 2004, 2005; Neuh\"auser et al. 2005; Caballero et
al. 2006b) and from exoplanets indirectly detected via radial velocity, transit
and microlensing methods (e.g. Mayor \& Queloz 1995; Charbonneau et al. 2000;
Beaulieu et al. 2006). 
The first spectroscopic data on IPMOs obtained by Zapatero Osorio et al. (2000)
confirmed the cool atmospheres of several of these objects, which appear rather
similar to those of PMOs orbiting stars.
The origin of both IPMOs and PMOs around stars remains uncertain.
It is likely that IPMOs form as a natural extension of the process that
leads to the formation of low-mass stars and, probably, brown dwarfs, but they
could also originate in protoplanetary discs and be ejected through dynamical
interactions  (Boss 2000; Reipurth \& Clarke 2001; Bate, Bonnel \& Bromm~2002).

It has been postulated that there is a minimum mass for formation of objects via
fragmentation in molecular clouds, the so called opacity mass-limit (Rees 1976;
Silk 1977; Tohline 1980; Bate, Bonnel \& Bromm 2003).
This theoretical limit is expected to be in the range 0.010--0.005\,$M_\odot$. 
It is therefore crucial to extend current surveys and identify objects with
masses as low as possible.  
The behaviour of the mass function at such low masses will be valuable
to establish the formation mechanism of the~IPMOs.  

At higher masses than in the planetary domain, the mass spectrum gives important
hints on how stars and brown dwarfs form in a molecular cloud (e.g. Kroupa 2001).
Following the nomenclature by Scalo (1986), the mass spectrum can be
approximated by a power-law function $\Delta N / \Delta 
{\rm M} \approx a {\rm M}^{-\alpha}$ (where $\alpha = - \gamma$ and $\Delta N$ is
the number of objects in the mass interval $\Delta {\rm M}$).
The substellar mass function in the brown dwarf domain has been investigated in
several open clusters, from the very nearby relatively old Hyades cluster to
much younger and distant star-forming regions
(see e.g. Zapatero Osorio et al. 1997; Bouvier et al. 1998; Luhman 2000; Barrado
y Navascu\'es et al. 2004; Bihain et al. 2006 and references therein).   
These studies point out a decrease of the slope in the power law when lower
masses are investigated.
However, the deepest searches suggest that the mass spectrum is still
rising even below the deuterium burning mass limit (B\'ejar et al. 2001;
Muench et al. 2002; Greaves, Holland \& Pound 2003; Oasa et al. 2006;
Gonz\'alez-Garc\'{\i}a et al.~2006).   

The $\sigma$~Orionis cluster displays characteristics that are advantageous over
other locations for the search and characterisation of substellar objects.  
In particular, it is nearby and very young.
Here we adopt an age of 3$\pm$2\,Ma (Oliveira et al. 2002; Zapatero Osorio et
al. 2002a; Sherry, Walter \& Wolk 2004) and a heliocentric distance
of 360$^{+70}_{-60}$\,pc (Brown, de Geus \& de Zeeuw 1994).
The cluster is relatively free of extinction ($A_V <$ 1\,mag; Lee 1968;
B\'ejar, Zapatero Osorio \& Rebolo 2004) and has a moderate spatial member
density, solar composition ([Fe/H] = +0.0$\pm$0.1; Caballero
2005) and a large frequency of intermediate-mass stars with discs
(33$\pm$6\,\%; Oliveira et al. 2006).
A compilation of different determinations of the age, distance and disc
frequency at different mass intervals is provided in Caballero (2007). 
The cluster contains several dozen brown dwarfs with spectroscopic features of
youth and with discs (Zapatero Osorio et al. 2002a; Barrado y Navascu\'es et al.
2002a; Muzerolle et al. 2003; Kenyon et al. 2005; Caballero et al. 2006a). 
It is also the star forming region with the largest number of detected candidate
IPMOs with follow-up spectroscopy (Zapatero Osorio et al. 2000, 2002b,
2002c; Mart\'{\i}n et al. 2001; Barrado y Navascu\'es et al. 2001; 
Mart\'{\i}n \& Zapatero Osorio~2003). 

Our present study is a natural extension of the work by B\'ejar et al.
(2001), who presented the first substellar mass function in the $\sigma$~Orionis
cluster. 
Here, we investigate the mass function down to a few Jupiter masses and use
recent data from 1 to 8\,$\mu$m that provide information on the existence of
circumsubstellar~discs.

\section{The optical-near infrared search}

%______________________________________________ f_m01
\begin{figure}
\centering
\includegraphics[width=0.51\textwidth]{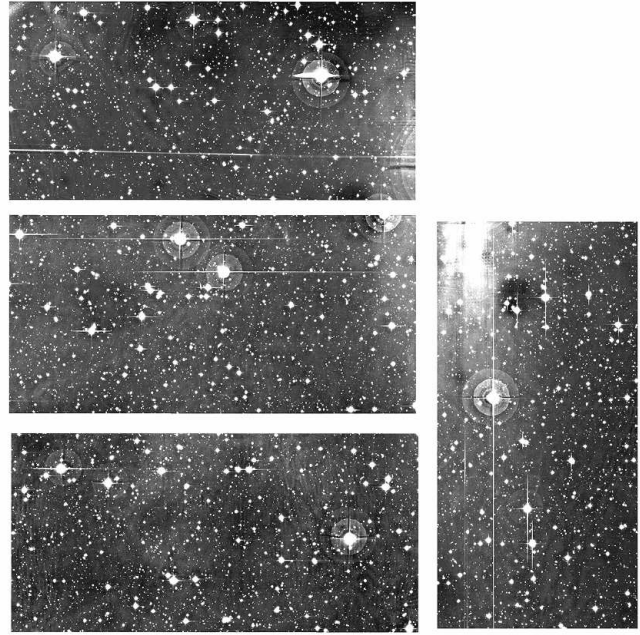}
\caption{Mosaic of the $I$-band images taken with WFC/INT (WFC03).
North is up and east to the left.
The size of each WFC chip is about 11 $\times$ 22 arcmin$^2$.
The glare of the OB star system $\sigma$ Ori is clearly visible to the northeast
of the vertical detector (CCD\#2). 
The mosaic on to a DSS-2-IR image centred on $\sigma$ Ori can be found in Fig. 4
in Caballero (2005).
See also the false-colour composite image in Fig.~\ref{ds9}, available
on-line.} 
\label{f_m01}
\end{figure}

We used the Wide Field Camera (WFC) at the 2.5-m Isaac Newton Telescope (INT)
and the Hawaii arm (SW) of ISAAC at the 8.2-m Very Large Telescope (VLT) UT1 
{\em Antu} to obtain very deep $I$- and $J$-band imaging. 
We studied a 0.22-deg$^2$ region to the southeast of the OB quintuple star
system \object{$\sigma$ Ori} that gives the name to the $\sigma$~Orionis
cluster. 
Table~\ref{log} lists the basic data of both optical and near-infrared
campaigns (dates, exposure times, survey area and average~seeing). 

%__________________________________________________ One column table
   \begin{table*}
      \caption[]{Basic log of the $I$- and $J$-band runs.} 
         \label{log}
     $$ 
         \begin{tabular}{c c c c c c c c c}
            \hline
            \hline
            \noalign{\smallskip}
Telescope	& 
Instrument	& 
Filter		& 
Dates 		& 
Exp. time 	&
Area		&
FWHM		&
Compl.		&
Limit.
\\
		&		
		&		
		&			
		& 
(s) 		&
(arcmin$^2$)	&
(arcsec)	&
(mag)		&
(mag)
\\
            \noalign{\smallskip}
            \hline
            \noalign{\smallskip}
INT		& 
WFC		& 
RGO $I$		& 
2000 Dec 30--2001 Jan 1 (WFC00) & 
21 $\times$ 1\,500		&
970		&
1.2--1.3	&
23.4		&
24.0
\\ % 5+10+6 repeticiones 		
		& 
		& 
		& 
2003 Jan 8 (WFC03) & 
20 $\times$ 1\,200 &
970		&
1.0--1.1	&
23.3		&
24.1
\\ % 12 repeticiones 		
{\em Antu} 	& 
ISAAC		& 
$J$		& 
2001 Dec 8,9,10,18 & 
4 $\times$ 60 	&
680		&
0.4--0.7	&
20.6		&
21.8	
\\ % 7$+$10$+$10$+$2 bloques; 1 bloque = 2 x 16 x 60 s
	 	& 
		& 
		& 
2001 Dec 8,9,10,18 & 
2 $\times$ 60$^{a}$ 	&
110		&
0.4--0.7	&
20.2		&
21.4	
\\ % 7$+$10$+$10$+$2 bloques; 1 bloque = 2 x 16 x 60 s
           \noalign{\smallskip}
            \hline
         \end{tabular}
     $$ 
\begin{list}{}{}
\item[$^{a}$] Areas along the borders of the ISAAC scans with shallower
integration depth.
\end{list}
   \end{table*}

The central coordinates of the survey (05:39:39 --02:44:40 J2000)
were chosen to take advantage of the asymmetric configuration of the four WFC
chips.   
Some of the brightest cluster stars, especially those of the central $\sigma$
Ori system and \object{HD 37525} AB, fell out of the field of view or in the
gaps between detectors.  
With this configuration, we found a compromise to study both the maximum area
close to the cluster centre, where the spatial density of cluster members is
larger, and the minimum area affected by the glare of the central star system
(see Fig.~\ref{f_m01}). 
The survey area is quite far from the location of the bulk of objects of the
group 1 in Jeffries et al. (2006), which have an average radial velocity
different from that adopted for the $\sigma$~Orionis cluster (see also
Caballero~2007).

\subsection{$I$-band data}

Two different optical datasets were obtained with the WFC/INT
(4 $\times$ 4k $\times$ 2k, 0.333\,arcsec\,pix$^{-1}$) and the RGO $I$ filter
(see Table~\ref{log}). 
Both datasets, named WFC00 and WFC03 after the years of observation (2000 and
2003), roughly share the same depth and coordinates of the centre of field.  
The older dataset (WFC00), however, was taken with a 25\,\% 
longer exposure time, leading to strong artefacts surrounding bright stars.
The seeing conditions were also slightly worse, which made the combination
of both datasets impractical. 
The WFC03 survey supersedes WFC00.
The night of 2003 Jan. 8 (WFC03) was photometric, allowing to image several
Landolt (1992) standard stars in the $I$-band.  

We reduced the data using standard procedures including bias subtraction and
flat fielding, and performed aperture and PSF photometry using {\tt noao.imred}
and {\tt noao.digiphot.daophot} routines within the IRAF environment, as
described in Caballero et al. (2004).  
About 30\,000 optical sources were detected in each dataset.
To avoid spurious detections in subsequent steps of the analysis, we
discarded sources with very large errors in their instrumental
magnitudes (roughly those with errors larger than 0.2 and 0.5\,mag for $I <$ 20
and $I>$ 20\,mag).
Thus there remained $\sim$25\,000 reliable optical sources for each~dataset.

The total covered field of view for each dataset was 970\,arcmin$^2$.
However, 3--4\,\% (WFC00) and 1--2\,\% (WFC03) of the area was useless due to
bleeding lines by saturated bright stars, their surrounding glares or
incomplete correction of chip dead columns during flat~fielding. 

Both datasets were also used to study the photometric variability of S\,Ori~45
(Zapatero Osorio et al. 2003) and another 27 young brown dwarfs of the $\sigma$
Orionis cluster (Caballero et al.~2004).

\subsection{$J$-band data}

We obtained 27 data blocks with ISAAC/{\em Antu} (1k $\times$ 2k,
0.148\,arcsec\,pix$^{-1}$) and the $J$ filter during three consecutive nights in
Dec. 2001.  
Two more data blocks were obtained in service mode one week later (2001 Dec
18; three $J$-band data blocks were missing to cover the $I$-band
survey completely). 
The sky was free of clouds except for the first night, that was partially
covered by thin cirrus.
The seeing was excellent during all the nights, with the FWHM measured in some
images below 0.5\,arcsec.  
Dark- and sky-subtraction, flat fielding and aligment and
combination of the data blocks were performed with IRAF. 
The overlapping area between optical (WFC) and near-infrared (ISAAC) images was
790\,arcmin$^2$. 

For easier handling, the individual $J$-band frames were aligned and
combined in long strips or scans, with sizes up to 22.0
$\times$\,2.4\,arcmin$^2$.  
Every region in the survey area except along the borders of the scans was
observed during 240\,s.  
The borders (1.2 $\times$ 2.4\,arcmin$^2$ on each side) were
exposed only half of the time of the main survey area.

Near-infrared point sources were detected using {\tt
noao.digiphot.daophot.daofind}. 
Aperture and PSF photometries were performed using other routines of the {\tt
noao.digiphot.daophot} package within the IRAF environment, in the same way as
in the optical.
We detected 11\,500 $J$-band sources.

\subsection{Photometric calibration and completeness}
\label{completeness}

We calibrated our $J$-band photometry to the Two-Micron All Sky Survey
Catalogue system (2MASS; Cutri et al. 2003) using selected sources in common
between our ISAAC images and the 2MASS catalogue. 
The useful overlapping magnitude interval between our deep images and the
2MASS data lies in the interval 14.5\,mag $\le J \le$ 15.5\,mag.
Hence, we could only use about a dozen comparison stars in each ISAAC scan to
establish the zero-point calibration for our near-infrared data.
This fact in turn led to calibration uncertainties twice larger than typical
2MASS errors. 
The standard deviation in the $J$ calibrations was in general less
than 0.09\,mag.   

The photometric calibration in the optical using photometric standard
stars from Landolt (1992) observed at different zenith distances during the
WFC03 run provided accuracies better than 0.05\,mag.
We only observed in the $I$-band filter, so no colour-dependent 
term was used in the photometric calibration.
As the Landolt standards are not very red, the photometry for the fainter
(redder) objects might suffer from unknown systematic effects.
Each of the four WFC chips was calibrated independently.
The WFC00 dataset was calibrated using bright WFC03 sources in common as a
reference.  
Except for intrinsically variable cluster members and for very faint optical
sources with large Poisson photon errors, the agreement between the
photometry of both WFC00 and WFC03 datasets is of the order of the
photometric calibration uncertainty (see further details in Caballero et al.
2004).  
We also compared the photometric data of objects with magnitudes fainter
than $I$ = 16.5\,mag in common with other independent surveys in the area (the
photometry of brighter ones may be affected by non-linear or saturation effects
in our WFC images). 
Within the uncertainties, there is no appreciable difference between the
$I$-band magnitudes in our study and those of B\'ejar et al. (1999, 2004).
However, Kenyon et al. (2005), based on Sloan Gunn $i'$- and Harris $R$-band
photometry, provided $I$-band magnitudes for objects in common with our survey
that are almost 0.3\,mag fainter.
Thus this difference in magnitudes is probably due to the different filter
systems used and/or the  absence of colour-dependent terms in our
photometric calibration.
The photometry from B\'ejar et al. (2001) must not be used for comparison, since
it has been found to be incorrect.  
The photometry from this work has been re-calibrated, using data taken the same
night as the one we are using to calibrate our $I$-band survey, and some
systematic errors due to variation of the zero-points between the WFC chips have
been found (V.~J.~S.~B\'ejar et al., in prep.).   

Completeness and limiting magnitudes in the $I$- and $J$-bands are shown in the
last columns of Table~\ref{log}.  
As completeness we take the point at which the number of detected sources per
magnitude interval stops increasing with a fixed power law of the magnitude,
$N(m) \propto m^p$ (where $p$ is an arbitrary real number). 
The measured number of sources departures 50\,\% from the power law at the
limiting magnitude. 
Our completeness and limiting magnitudes are roughly equivalent to the
10\,$\sigma$ and 3\,$\sigma$ magnitudes, respectively.
The shallower borders along the ISAAC scans led to brighter
completeness and detection limits by $2.5 \log{2^{1/2}} \sim$ 0.4\,mag. 

Given the completeness limits of the $I$ and $J$ surveys and the expected
colours of the objects of interest, the $I$-band photometry is constraining
the survey at the fainter end.
Our faintest cluster member candidates are however brighter than the
completeness limits in $I$ and $J$ (see Section~\ref{selection}).
The search is also limited in the brightest magnitude interval by the optical
data. 
The confirmed stellar cluster member \object{S\,Ori~8}, with $J$ =
14.14$\pm$0.03\,mag and $I$ = 15.74$\pm$0.04\,mag (2MASS Catalogue and B\'ejar
et al. 1999), was detected in the non-linear regime of the near-infrared images,
but saturated in the optical images. 
It is the only known cluster member in the area fainter than $J$ =
14.0\,mag that has not been studied here.

\subsection{Astrometric calibration and combination of $I$- and $J$-band data}

First, we transformed the physical coordinates of the optical sources on
the WFC chips and of the near-infrared sources on the ISAAC scans to celestial
coordinates using the USNO-A2 and 2MASS astro-photometric catalogues as
references (see further details in Bihain et al. 2006).
The accuracies in the astrometric solutions were about 0.1 and 0.2\,arcsec
for the $I$-band images and the $J$-band scans, respectively (about twice the
average catalogue errors).
 
After the positional cross match between near-infrared and optical sources, we
got the mean coordinates and $I$- and $J$-band magnitudes for $\sim$9400
objects. 
About 2100 near-infrared sources were not correlated with any optical source. 
Among them, a few hundred objects within the completeness of the $J$-band images
and without optical counterparts (i.e. with expected red $I-J$ colours) were
visually inspected on the WFC images, since they could be very low-mass
cluster members fainter than the completeness of the optical data. 
After this analysis, we concluded that the vast majority were not detected in
the $I$-band images because they were: ($i$) double or multiple background stars
or galactic nuclei only resolved in the ISAAC data, ($ii$) faint sources in the 
glare of bright stars in the optical, ($iii$) artefacts in the ISAAC frames,
($iv$) multiple detections of saturated bright stars or ($v$) extended sources
of extragalactic nature.
Some promising near-infrared sources without optical counterpart were
followed up with deep $HK$ imaging (see Section \ref{followup}).

\section{Cluster member selection and the infrared follow-up}

%__________________________________________________ spektra
   \begin{table*}
      \caption[]{Objects with signatures of extreme youth and cluster membership
	in the survey area compiled from the literature.}   
         \label{spektra}
     $$ 
         \begin{tabular}{l c c c c c l}
            \hline
            \hline
            \noalign{\smallskip}
Name				& pEW(Li {\sc i})	& pEW(H$\alpha$)& low $g^{a}$ 	& V$_r^{b}$	& Remarks & References 	\\  
				& (\AA)			& (\AA) 	& 		& 	& 	& 					\\  
             \noalign{\smallskip}
            \hline
            \noalign{\smallskip}
S\,Ori~J054000.2--025159	& +0.25$\pm$0.09 	&  		& yes 		& bin.	&	& Ke05 					\\ % 11142
S\,Ori~J053848.1--024401	& +0.53$\pm$0.10 	&  		& yes 		& yes	&	& Ke05 					\\ % 21184
S\,Ori~J053833.9--024508	& +0.38$\pm$0.07 	&  		& yes 		& yes	&	& Ke05 					\\ % 22241
S\,Ori~J053911.4--023333	& +0.33$\pm$0.06 	& --4.7$\pm$1.0 & yes 		& yes	&	& ByN03, Ke05 				\\ % 33098
S\,Ori~14			& +0.71$\pm$0.06 	&  		& yes 		& yes	&	& Ke05 					\\ % 42849
S\,Ori~J054014.0--023127	& +0.47$\pm$0.08 	& broad 	& yes 		& yes	& var.	& Ca04, Ke05 				\\ % 32232
S\,Ori~J053847.2--025756	& +0.53$\pm$0.10 	& broad 	& yes 		& yes	& var.	& Ca04, Ke05 				\\ % 21798
S\,Ori~J053838.6--024157	& +0.55$\pm$0.06 	& --6$\pm$1 	& yes 		& yes	&	& Ke05, Bu05, Ca06 		\\ % 22097
S\,Ori~25			& +0.53$\pm$0.10 	& --45.0$\pm$1.0& yes 		& yes	& var., He {\sc i}, X	& B\'e99, Mu03, Fr06 	\\ % 41080
S\,Ori~J053826.1--024041	& +0.51$\pm$0.08 	& --4$\pm$2	& yes 		& yes	&	& ByN03, Ke05 				\\ % 23044
S\,Ori~27			& +0.74$\pm$0.09 	& --5.7$\pm$1.5	& yes 		& yes	&	& ZO02, Ke05 				\\ % 24047
S\,Ori~J053825.4--024241	&  			& --260$\pm$30	&  		& yes	& var., nIR, He {\sc i}	& Ca06 			\\ % 23156
S\,Ori~28			& +0.66$\pm$0.09 	& 		& yes 		& yes	&	& Ke05 					\\ % 43752
S\,Ori~32			& +0.46$\pm$0.09 	& 		& yes 		& yes	&	& Ke05 					\\ % 44072  	 
S\,Ori~J054004.5--023642	& $<$0.39 		& broad		& yes 		& yes	& var. & Ke05 				\\ % 34252  	    
S\,Ori~36			& +0.45$\pm$0.15 	& 		& yes 		& bin.	&	& Ke05 					\\ % 34332
S\,Ori~42			&  			& --89$\pm$12	&  		& 	& var.	& ByN03, Ca04 				\\ % 41389
S\,Ori~45			& +2.4$\pm$1.0 		& [--60,--20]	& yes 		& bin.?	&	& B\'e99, ZO02 				\\ % 23438
S\,Ori~51			&  			& --25:		& yes($J$) 	& 	&	& ByN03, McG04 				\\ % 31624
S\,Ori~71			&  			& --700$\pm$80	&  		& 	& var.?	& Ca04, ByN02 				\\ % 34380
\noalign{\smallskip}
            \hline
         \end{tabular}
     $$ 
\begin{list}{}{}
\item[$^{a}$] Low-gravity features.
\item[$^{b}$] Radial velocity consistent with membership in cluster.
Symbol `bin.' denotes possible binary according to Kenyon et al.
(2005).
\end{list}
   \end{table*}

\subsection{Known objects in the survey area}

In the studied area and magnitude intervals (16.1\,mag\,$\lesssim I
\lesssim$\,24.1\,mag and 14.2\,mag\,$\lesssim J \lesssim$\,20.6\,mag) there are
50 cluster members and cluster member candidates reported in the
literature. 
They have been studied in the optical surveys by: 
B\'ejar et al. (1999 --B\'e99--, 2001, 2004); Zapatero Osorio et al.
(2000); Caballero et al. (2004 --Ca04--); Kenyon et al. (2005 
--Ke05--); Gonz\'alez-Garc\'{\i}a et al. (2006),   
and in specific works such as:  
Zapatero Osorio et al. (1999, 2002a --ZO02--); Barrado y Navascu\'es et
al. (2002a --ByN02--, 2003 --ByN03--); Muzerolle et al. (2003 
--Mu03--); McGovern et al. (2004 --McG04--); Scholz \& Eisl\"offel
(2004); Burningham et al. (2005a --Bu05--); Caballero et al. (2006a 
--Ca06--); Franciosini, Pallavicini \& Sanz-Forcada (2006 --Fr06--).

Out of these 50 objects, only one, \object{S\,Ori~69}, was not detected on
our images.  
This non-detection suggests that the isolated planetary-mass object
candidate is fainter than what was published in the literature
($I = 23.9\pm0.2$\,mag, $J = 20.2\pm0.4$\,mag; Zapatero Osorio et al. 2000;
Mart\'{\i}n et al. 2001).
G.~Bihain et al. (private comm.) have recently obtained new deep
near-infrared imaging of S\,Ori~69, determining its $J$-band magnitude at 21.61
$\pm$ 0.16\,mag.  
Two out of the 49 identified targets in the area, \object{S\,Ori~41} and
\object{S\,Ori~J054004.9--024656}, were classified as probable non-members by
B\'ejar et al. (2001, 2004) (the latter is a visual binary resolved in our ISAAC
images, with $\rho \lesssim$ 0.8\,arcsec).  
Also, the M5.0$\pm$0.5-type dwarf \object{S\,Ori~J053909.9--022814} does not
display signatures of youth in high-quality optical spectra (Barrado y
Navascu\'es et al. 2003; Kenyon et al. 2005). 
We do not consider these three objects as cluster members.

There remain 46 cluster members and candidates reported in the literature in the
survey area. 
Spectroscopic information is available for 31 of them.
For 11 of these objects, only spectral types could be determined, while the
other 20 members display spectroscopic features of extreme youth (age $<$
10\,Ma). 
The considered features are: Li~{\sc i} $\lambda$6707.8\,\AA~in absorption,
broad and/or strong H$\alpha$ emission, weak alkali absorption lines (i.e.
low-gravity; from the pEW(Na~{\sc i}) in optical spectra, except in the case of
S\,Ori~51 --from a $J$-band spectrum--), and emission lines ascribed to
accretion processes or outflows (e.g. [O~{\sc i}] $\lambda$6300.3\,\AA,
He~{\sc i} $\lambda$5875.8\,\AA).   
Most of them have radial velocities, V$_r$, similar to the cluster systemic
radial velocity at about 30--35\,km\,s$^{-1}$.
Additionally, some of them display optical photometric variability with
amplitude larger than 0.07\,mag (`var.'), $K_{\rm s}$-band flux excess
(`nIR'), or X-ray emission (`X').
There is insufficient and/or discrepant membership information on S\,Ori~20
(Barrado y Navascu\'es et al. 2003; Kenyon et al. 2005), S\,Ori~47 (Zapatero
Osorio et al. 1999; McGovern et al. 2004), and S\,Ori~J053844.4--024037
(Burningham et al. 2005a).  
We do not consider these three objects as spectroscopically {\em confirmed}
young cluster members.
The 20 confirmed cluster members together with their references and most
important characteristics are given in Table~\ref{spektra}.

\subsection{Cluster member selection from the $I$ vs. $I-J$ diagram} 
\label{selection}

%______________________________________________ f_IvsIJ
\begin{figure}
\centering
\includegraphics[width=0.49\textwidth]{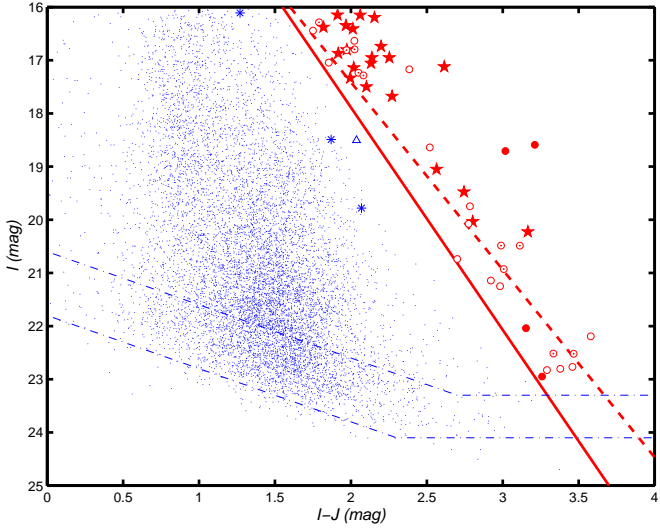}
\caption{$I$ vs. $I-J$ colour-magnitude diagram from our WFC03/INT-ISAAC/{\em
Antu} survey.
The different symbols indicate:
probable fore- and background sources (small dots, `$\cdot$');
confirmed cluster members (filled stars, `$\star$');
cluster member candidates with spectral type determination (encircled
dots, `$\odot$'); 
cluster member candidates without spectral type determination (open circles,
`$\circ$'); 
the four new cluster member candidates (filled circles, `$\bullet$');
S\,Ori~J053948.1--022914 (the open triangle, `$\triangle$');
the three detected cluster non-members (asterisks, `*').
Dash-dotted lines indicate the completeness and detection limits of our survey.
The dashed line is the lower envelope of confirmed cluster members with
youth indicators.
The solid line is the boundary considered.
Colour versions of all our Figures are available in the electronic
publication.}  
\label{f_IvsIJ}
\end{figure}

\begin{figure}
\centering
\includegraphics[width=0.49\textwidth]{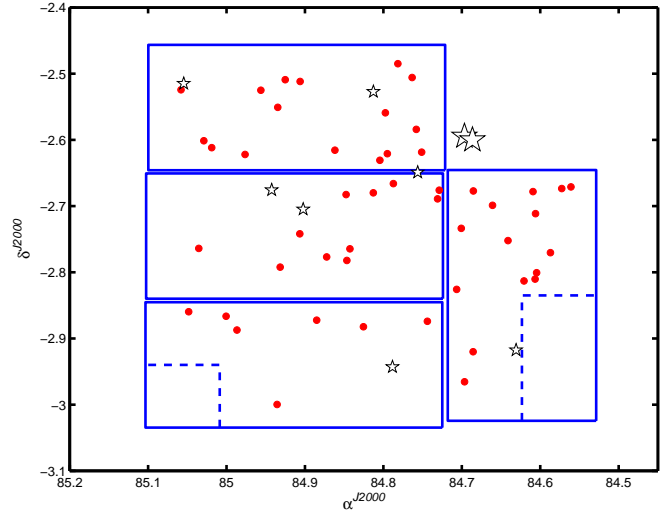}
\caption{Pictogramme showing our $IJ$ search.
The incomplete coverage of the $J$ survey is indicated with dashed lines. 
Code: filled circles, our cluster members and candidates;
big open stars, $\sigma$~Ori~AB and E (at the centre of the cluster); 
small open stars, other young stars in the survey area more massive than
1.2\,$M_\odot$ from Caballero (2007).
Note the increasing density of objects towards the centre of the cluster.}  
\label{mayrit_apuntados}
\end{figure}

We have selected the cluster member candidates from our WFC03 + ISAAC data,
using the $I$ vs. $I-J$ colour-magnitude diagram shown in Fig.~\ref{f_IvsIJ}. 
Among the 9400 sources with optical and near-infrared information in our
survey, we have selected new photometric candidate cluster members based on
their position in the diagram with respect to the 20
spectroscopically confirmed cluster members, indicated by filled stars.
The latter define the spectro-photometric sequence of the $\sigma$~Orionis
cluster, that is redder than the location of field stars.
Most of them are brighter than $I$ = 18.0\,mag, while only four brown
dwarfs with spectroscopic features indicative of youth fall in the interval 19.0
$\le I \le$ 20.5\,mag. 
Five brown dwarf and IPMO candidates fainter than $I$ = 19.0\,mag
have low-resolution spectroscopy available in the literature and
extrapolate the spectro-photometric cluster sequence towards fainter 
magnitudes and lower masses.
The lower envelope of the confirmed candidates from $I$ = 16.0 to 20.5\,mag is
indicated by a dashed line in the diagram.
Our adopted selection criterion for cluster membership ({\em solid} line in
Fig~\ref{f_IvsIJ}) is the lower envelope of cluster members ({\em dashed} line)
shifted to the blue to accommodate photometric uncertainties. 
To account for increasing photometric errors at fainter magnitudes, this
envelope is displaced by 0.05\,mag at $I$ = 16\,mag and 0.45\,mag at $I$ =
25\,mag.
There are 49 objects (45 previously known) to the right of this boundary, which
is the final cluster  list we will use throughout the present paper.
All previously reported $\sigma$~Orionis candidates, located within the survey 
area and magnitude limits, lie to the right of the membership boundary -- with
the exception of \object{S\,Ori~J053948.1--022914}, whose properties are
summarised in Section~\ref{022914}.
Identifications, J2000 coordinates, WFC03/INT
$I$-band and ISAAC/{\em Antu} $J$-band magnitudes and spectral types when
available of the 49 selected targets are detailed in Table~\ref{pre}.  
Confirmed cluster members with at least one feature of youth are also indicated
with a ``Y''.
The spatial distribution of the 49 $\sigma$~Orionis members and member
candidates is shown in Fig.~\ref{mayrit_apuntados}.

Of the list of 49 objects, there are four identified for the first time in
our survey.
They are marked with the label ``New'' in Table~\ref{pre}.
For their identifications, we follow the series started by B\'ejar et al.
(2001), who used the nomenclature S\,Ori~Jhhmmss.s--ddmmss for their
candidate $\sigma$~Orionis cluster members.
The ``S\,Ori'' sources are, however, not associated with the variable star
\object{S~Ori} (HD~36090), which is located several degrees away in the
constellation of~Orion.
The two brightest new objects are red candidate brown dwarfs that were embedded
in the glare of $\sigma$~Ori in the northwest corner of CCD\#4 (in the centre of
the WFC00 mosaic) in Caballero et al. (2004). 
The other two new objects, with $I$ = 22--23\,mag, fell in the $I$ vs. $I-J$
diagram slightly to the left of the 10\,Ma-old Dusty00 isochrone (Chabrier et
al. 2000), which was used by Caballero et al. (2004) to differentiate candidate
cluster members from probable fore- and back-ground sources.

Six bright sources in our final sample were saturated in the
longer-exposure WFC00 images, too.
All of our $\sigma$~Orionis members and member candidates have been 
selected from the combination of the WFC03 optical dataset and the ISAAC
$J$-band images.
The combination of the WFC00 optical dataset and the ISAAC data yields identical
results in the magnitude interval common to the two epochs of WFC observations.
The exceptions are the two candidate brown dwarfs whose photometry was
contaminated by the glare of $\sigma$~Ori in the WFC00 data and the six
stars saturated in the longer WFC00 exposures. 
This supports our selection of cluster candidates and provides evidence for the
fact that photometric optical variability does not significantly affect our
selection criterion.

%______________________________________________ mayrit_apuntados

\subsection{The infrared follow-up}

To discriminate foreground contaminants among the list of selected objects, we
have taken infrared data from the IRAC/{\em Spitzer} Space Telescope archive and
from 2MASS, and performed new deep $H$- and $K_{\rm s}$ imaging.

%__________________________________________________ One column table
   \begin{table*}
      \caption[]{Log of the $HK$-band follow-up campaigns.} 
         \label{log-nir}
     $$ 
         \begin{tabular}{l c c c c c c c c}
            \hline
            \hline
            \noalign{\smallskip}
Telescope	& 
Instrument	& 
Filter(s)	& 
Date(s) 	& 
Exp. time 	&
Area		&
FWHM		&
Compl.		&
Limit.
\\
		&		
		&		
		&			
		& 
(s) 		&
(arcmin$^2$)	&
(arcsec)	&
(mag)		&
(mag)
\\
            \noalign{\smallskip}
            \hline
            \noalign{\smallskip}
CFHT		& 
CFHT-IR 	& 
$H$, $K'$	& 
2004 Feb 9,17,22 & 
1800 		&
85		&
0.7--1.0	&
20.0--20.5	&
21.0--21.5	
\\ % 
3.5\,m CA	& 
$\Omega$2k	& 
$H$		& 
2003 Oct 19,22 	& 
1200--7200	&
$\sim$1100	&
$>$1.5		&
16.8		&
18.1
\\ % 		
		& 
		& 
$H$, $K_{\rm s}$ & 
2005 Jan 27--Feb 1 & 
5100--11700	&
480		&
1.3--2.0	&
17.7--19.5	&
18.5--20.0
\\ % 		
TCS		& 
CAIN-II		& 
$J,H,K_{\rm s}$ & 
2004 Apr 19     & 
3600--7200	&
$<$100		&
$\sim$1.2	&
18.5, 18.0, 17.5 &
19.5, 19.0, 18.5
\\ % 		
		& 
		& 
$K_{\rm s}$     & 
2006 Oct 20--22 & 
2400--3600	&
$<$100		&
$\sim$1.0	&
$\sim$17.5	&
$\sim$18.5
\\ % 		
           \noalign{\smallskip}
            \hline
         \end{tabular}
     $$ 
   \end{table*}

\subsubsection{IRAC/{\em Spitzer} follow-up}

We used the Infrared Array Camera (IRAC) deep image mosaics
centred on $\sigma$~Ori from the {\em Spitzer} Space Telescope Archive
at the 3.6, 4.5, 5.8, and 8.0\,$\mu$m bands (denoted here as $[3.6]$,
$[4.5]$, $[5.8]$, and $[8.0]$). 
We got the post-basic calibrated data with the tool {\em Leopard},
performed 5-arcsec standard aperture photometry with IRAF on sources with
S/N ratio $>$ 3, applied the corresponding aperture correction factors, and
converted the measured flux per pixel into calibrated magnitudes in the Vega
system using the zero-points for each channel of the IRAC instrument.
We performed additional PSF photometry of some of the faintest sources. 
Neither array-location-dependent nor colour corrections were applied.
Up-to-date information on the photometry and absolute calibration of IRAC data
is given at {\tt  http://ssc.spitzer.caltech.edu/irac}.
The completeness magnitudes of the mosaic images in the $[3.6]$, $[4.5]$,
$[5.8]$, and $[8.0]$ bands are 17.2, 17.2, 15.0, and 14.0\,mag, respectively,
with uncertainties of 0.3\,mag.
Our magnitudes deviate no more than 0.10\,mag from the independently
calibrated magnitudes of objects in common with Hern\'andez et al. (2007). 

Of the 49 objects in our sample, three fall out of the IRAC field of view
and two, the most distant objects to the cluster centre, are only in the $[4.5]$
and $[8.0]$ mosaics.
Table~\ref{nir} summarises the available IRAC data for 46 cluster members and
candidates.
Typical errors of the IRAC photometry vary between $\lesssim$ 0.1\,mag for the
brightest targets to $\lesssim$ 0.5\,mag for the faintest ones.
Non-detected objects are fainter than the completeness limit in each
band. 
In particular, 8 and 15 objects are too faint to be detected in the
$[5.8]$ and $[8.0]$ bands, respectively. 
The fourth digit in the Flag column indicates the number of IRAC channel mosaics
in which an object is located.

\subsubsection{Ground-based $HK$-band follow-up}
\label{followup}

2MASS $JHK_{\rm s}$ photometry is available for 33 targets (the 32 brightest
cluster member candidates, plus S\,Ori~47). 
Since ISAAC images were calibrated with 2MASS, the agreement between both
$J$-band magnitudes is excellent ($\Delta J = J({\rm ISAAC}) - J({\rm 2MASS}) =
-0.03 \pm 0.11$\,mag).
However, there are hints for the photometric variability of some targets from
the 2MASS epoch (MJD = 51\,116.3) to the ISAAC epoch (MJD $\approx$ 52\,250)
which require further studies.
Some of them are previously known variables like S\,Ori~42 (Caballero et al.
2004) and S\,Ori~J053825.4--024241 (Caballero et al. 2006a), or are found
in this work for the first time (e.g. S\,Ori~J053902.1--023501; $J_{\rm ISAAC} -
J_{\rm 2MASS}$ = 0.25$\pm$0.12\,mag).   

To complement our $J$-band data and the relatively-shallow $HK_{\rm s}$-band
data from 2MASS, we performed deep ground-based near-infrared imaging of
most of the faintest targets of our sample. 
The follow-up was performed using CFHT-IR at the 3.58-m Canada-France-Hawai'i
Telescope (1k $\times$ 1k; 0.211\,arcsec\,pix$^{-1}$), Omega-2000 ($\Omega$2k)
at the 3.50-m Calar Alto Teleskop (2k $\times$ 2k; 0.450\,arcsec\,pix$^{-1}$)
and CAIN-II at the 1.52-m Telescopio Carlos S\'anchez (256 $\times$ 256;
1.00\,arcsec\,pix$^{-1}$). 
Table~\ref{log-nir} summarises the follow-up campaigns. 
Each of the CFHT-IR and CAIN-II pointings imaged a single target, whereas
several targets were imaged simultaneously in the large field of view of
$\Omega$2k.  
The first run with $\Omega$2k, not as deep as the second, was performed only in
the $H$ band, but covering 95\,\% of the overlapping area between the WFC and
ISAAC images.
The pointings with CFHT-IR were devoted to the follow-up of four $J$-band
sources whose $I$-band counterpart was not automatically detected (see
Section~\ref{newinteresting}, where we present four interesting red objects that
probably do not belong to the $\sigma$~Orionis cluster). 

For CFHT-IR, we took 30 individual frames of 60\,s exposure time each per
pointing and per filter, using a regular dithering pattern with a suitable
shift.  
A bad-pixel mask and a dome flat-field image for each filter were used during
the reduction.
The $\Omega$2k observations were similar to those with CFHT-IR, but using
shorter exposure times (1.6\,s --$K_{\rm s}$--, 2 or 3\,s --$H$--) and a random
dithering pattern.   
For CAIN-II, the dithering pattern was of 10 positions and the individual
exposure times were of 10\,s ($H$) and 6\,s ($K_{\rm s}$). 
Photometric calibration and astrometry for CFHT-IR, $\Omega$2k and CAIN-II
were identical to those for the ISAAC $J$-band images. 

The available $JHK_{\rm s}$ magnitudes of the objects in our sample are
provided in Table~\ref{nir}.
Apart from the 2MASS catalogue and our ground-based follow-up, we have also
taken the $K_{\rm s}$ magnitude of three candidate cluster members from the
literature. 
The first three digits in the Flag column in the Table inform about the source
of the $J$-, $H$- and $K_{\rm s}$-band photometry
(0: no data available; 1: this paper; 2: 2MASS; 3: Mart\'{\i}n et al. 2001).
The uncertainty in our $HK_{\rm s}$ magnitudes are in general smaller than
0.1\,mag, and are larger than 0.2\,mag only in a very few cases.

\subsubsection{Blue interlopers and theoretical isochrones}

%______________________________________________ f_IvsIKs & f_JKsvsIJ
\begin{figure*}
\centering
\includegraphics[width=0.49\textwidth]{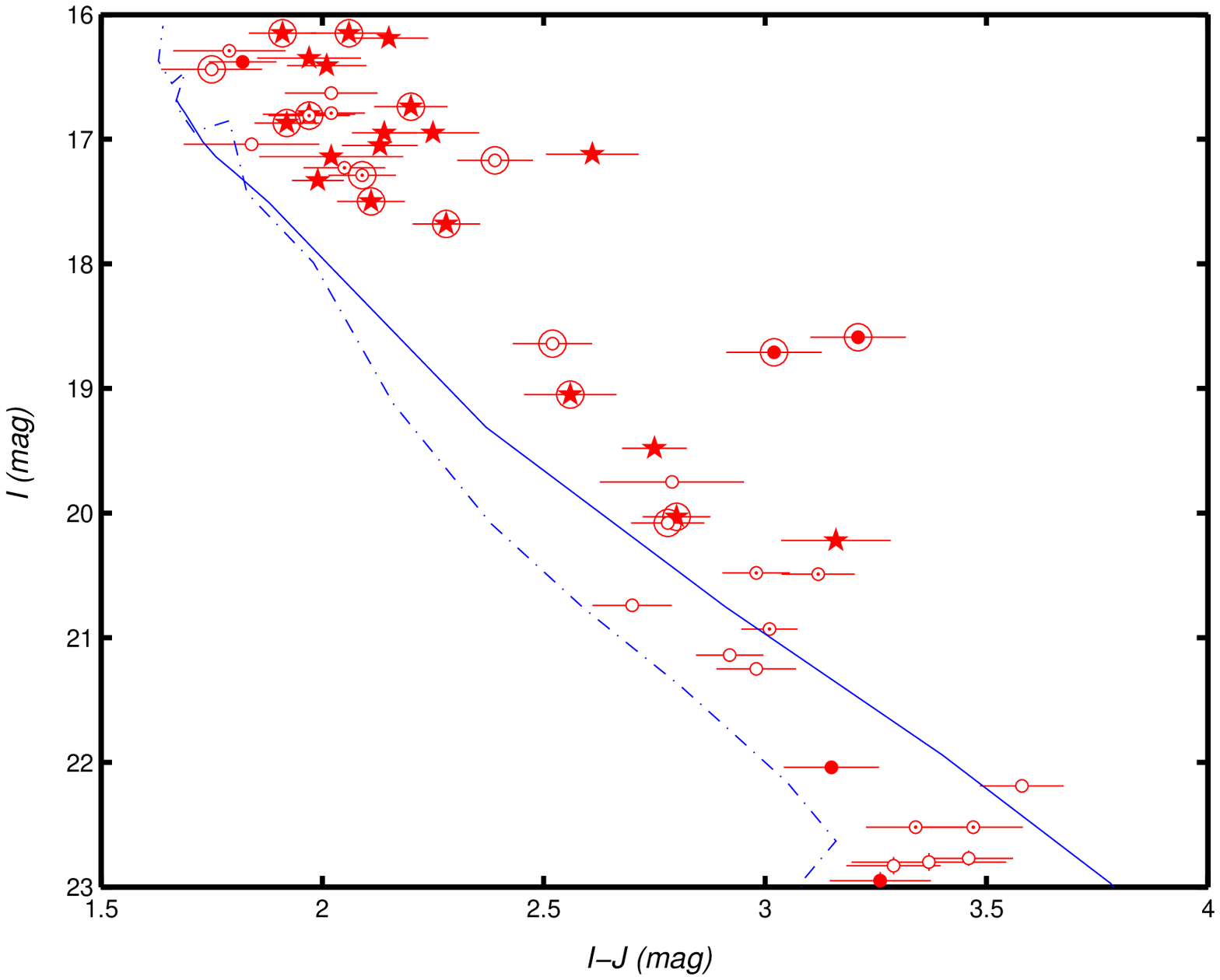}
\includegraphics[width=0.48\textwidth]{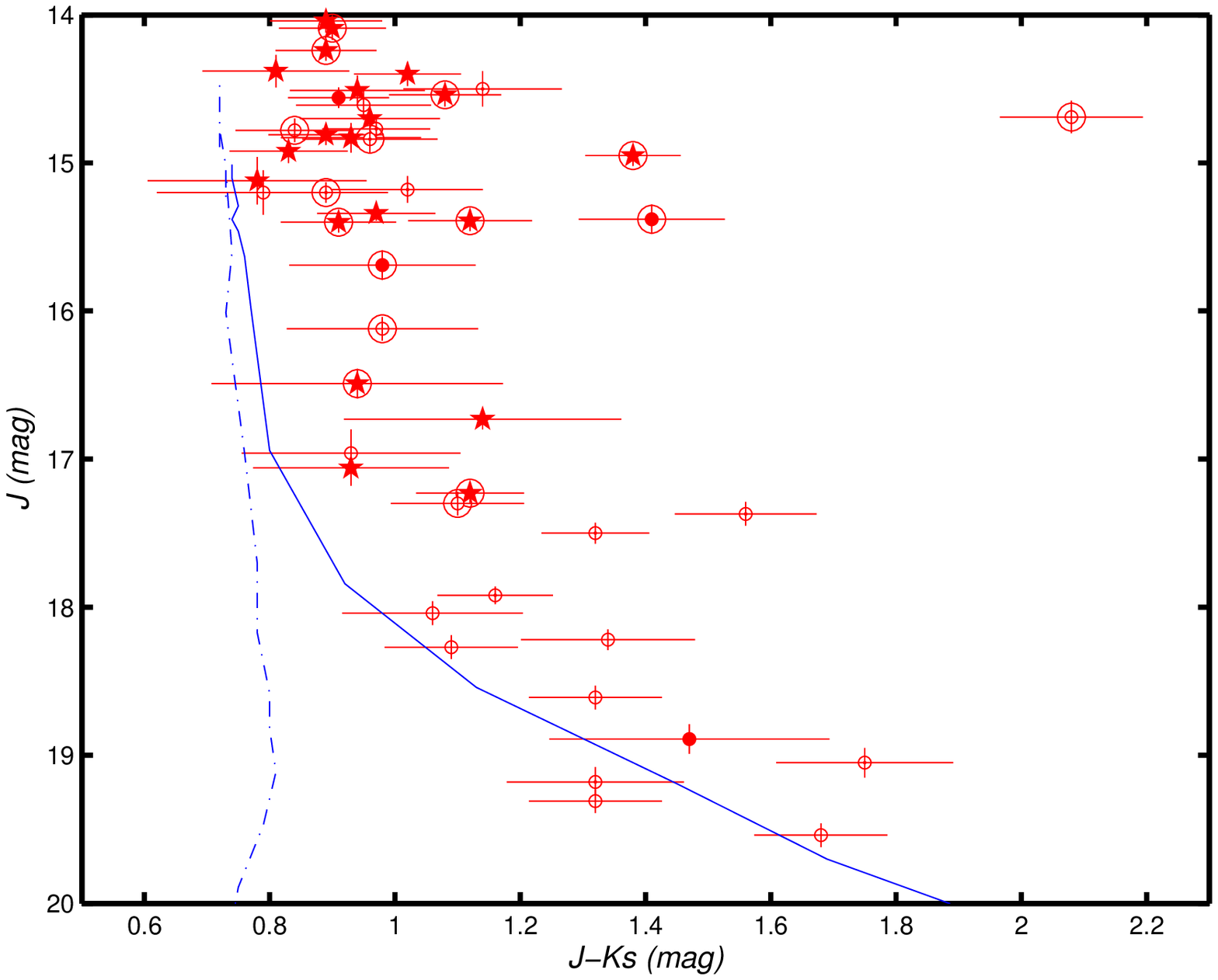}
\includegraphics[width=0.49\textwidth]{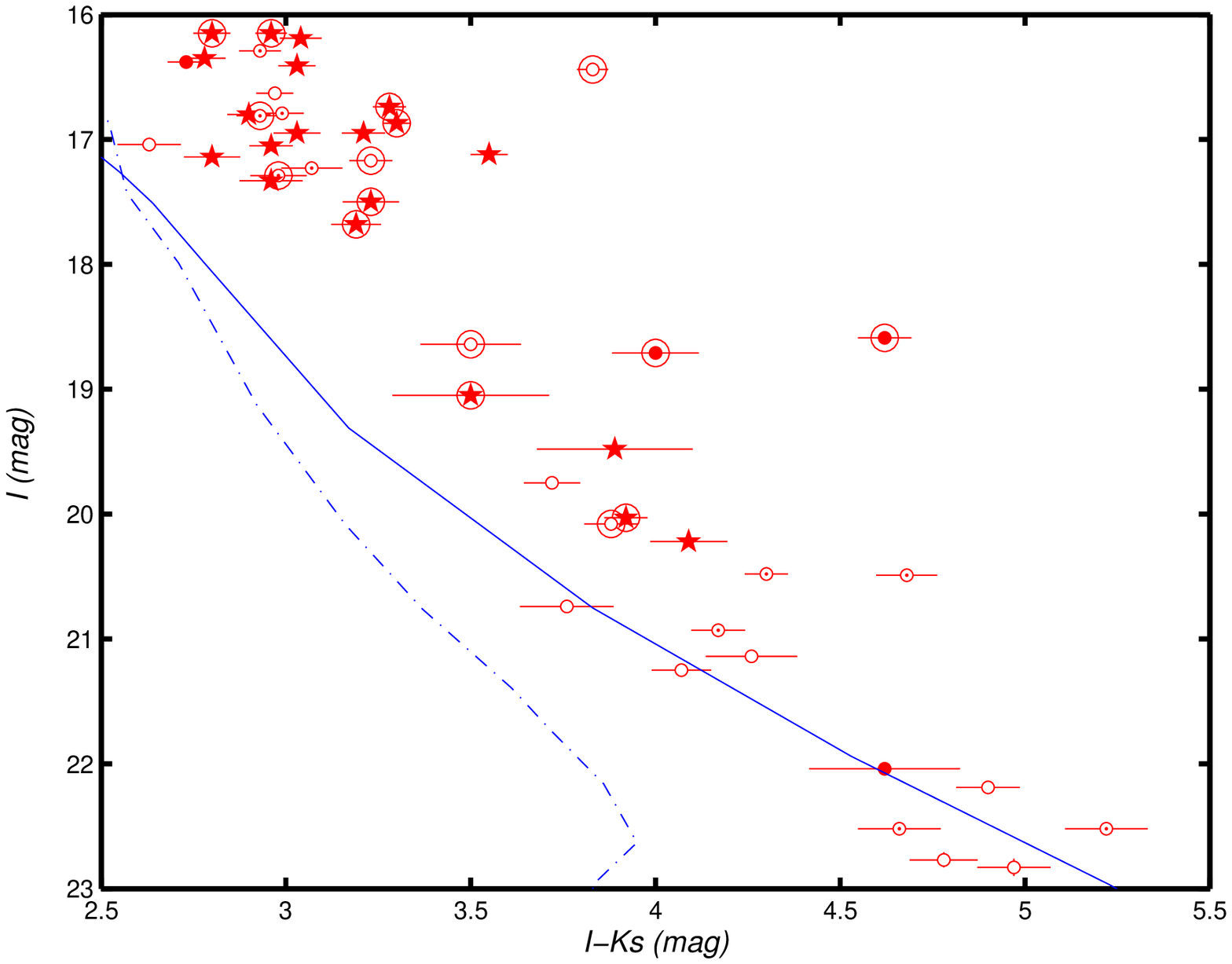}
\includegraphics[width=0.48\textwidth]{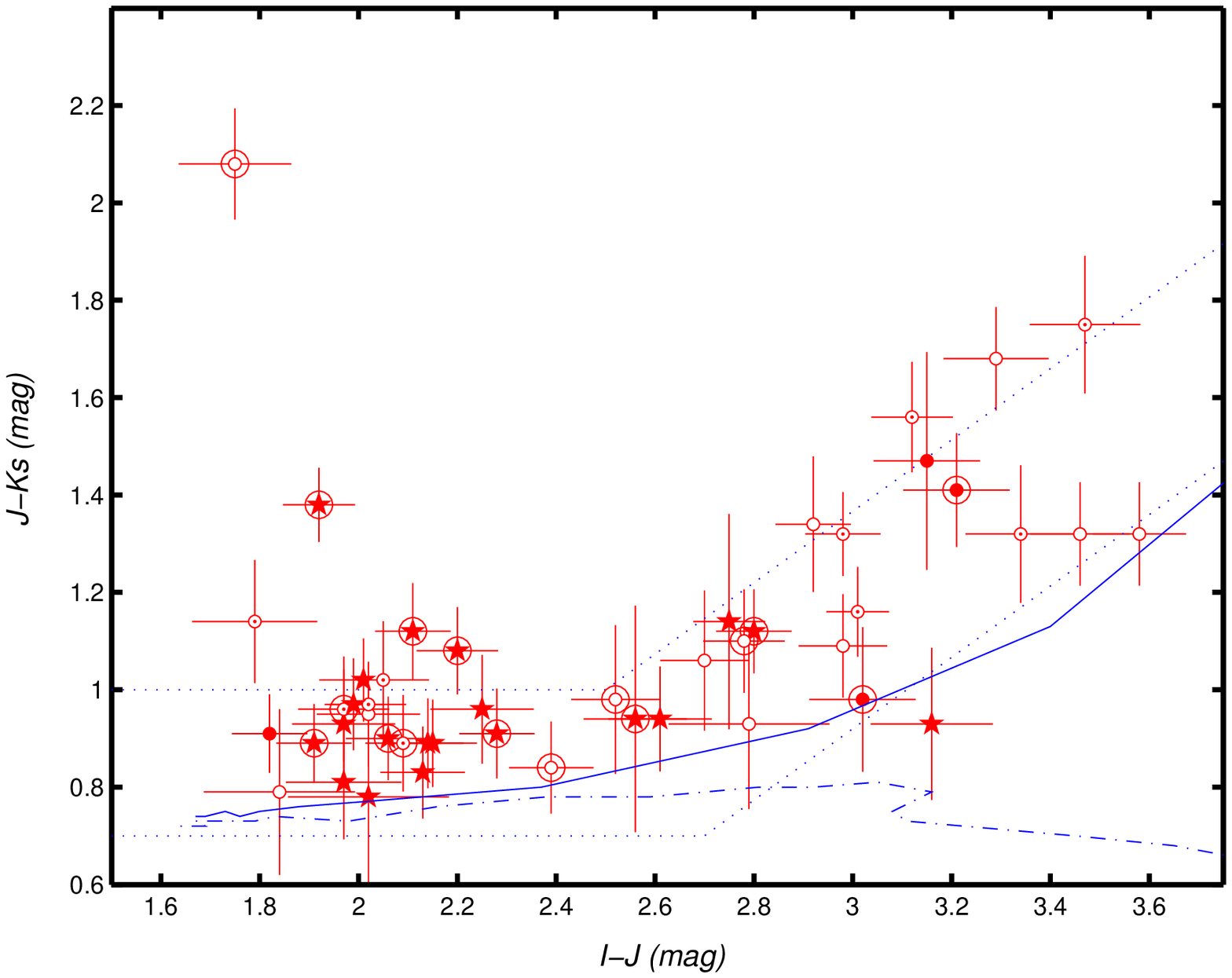}
\caption{$I$ vs. $I-J$ (top left), $I$ vs. $I-K_{\rm s}$ (bottom left), and $J$
vs. $J-K_{\rm s}$ (top right) colour-magnitude diagrams, and  $J-K_{\rm s}$ vs.
$I-J$ colour-colour diagram (bottom right).
$I$ is from WFC03, $J$ from ISAAC and $HK_{\rm s}$ from the ground-based
follow-up data.
The symbols are as in Fig.~\ref{f_IvsIJ}.
Objects with discs from IRAC excess flux are marked with big circles.
Solid and dash-dotted lines are the Dusty00 and Cond03 3\,Ma-old
isochrones at 360\,pc of the Lyon group. 
Dotted lines in the colour-colour diagram mark the approximate sequence of field
ultracool dwarfs (Dahn et al. 2002). 
The very red object with $I-J$ = 1.75$\pm$0.11\,mag and $J-K_{\rm s}$ = 
2.08$\pm$0.11\,mag is S\,Ori~J053902.1--023501, whose spectral energy
distribution (Fig.~\ref{f_seds}.) shows evidence for a circumstellar disc
(Section~\ref{section.discs}).}
\label{f_IvsIKs}
\end{figure*}

To sum up, we have collected near-infrared follow-up photometry for 47
selected cluster member candidates (30 targets with photometry in eight
passbands $IJHK_{\rm s}[3.5][4.5][5.8][8.0]$; 45 with photometry in at
least six passbands). 
We have plotted all the possible colour-magnitude diagrams, in search for ``blue
interlopers'', i.e. possible contaminants that satisfy the $IJ$ selection
criterion, but display bluer colours than expected for cluster members in other
diagrams. 
In Fig.~\ref{f_IvsIKs} we show three colour-magnitude diagrams of our sample
($I$ vs. $I-J$, $I$ vs. $I-K_{\rm s}$, $J$ vs. $J-K_{\rm s}$). 
We cannot determine if the two faint objects with only $I$- and $J$-band
photometry available, S\,Ori~J054008.5--024551 (previously classified as 
high-probability cluster member in the $i'z'$-band search with $J$-band
follow-up by Gonz\'alez-Garc\'{\i}a et al. 2006) and S\,Ori~J054011.6--025135,
are blue contaminants or not. 
Two possible blue-interloper candidates have been identified.
From the $J-K_{\rm s}$ vs. $I-J$ colour-colour diagram in the bottom right panel
of Fig.~\ref{f_IvsIKs}, S\,Ori~51 ($I-J$ = 3.16$\pm$0.13\,mag; $J-K_{\rm
s}$ = 0.93$\pm$0.16\,mag) displays a $J-K_{\rm s}$ colour that is marginally blue
if compared to those of other $\sigma$~Orionis cluster members and field
ultracool dwarfs with similar $I-J$ colour (e.g. Dahn et al. 2002). 
In contrast, its position matches the spectro-photometric cluster sequence in
all the colour-magnitude diagrams and its $J$-band spectrum displays low-gravity
features, which support cluster membership (McGovern et al. 2004). 
The second one, S\,Ori~J053922.2--024552 ($I$ = 17.04$\pm$0.03\,mag;
$I-K_{\rm s}$ = 2.63$\pm$0.09\,mag), without spectroscopic
information, has an $I-K_{\rm s}$ colour that is slightly bluer than the
confirmed cluster members of the same magnitude in our survey.
It has, however, similar colours and magnitudes to \object{SE 70}, a $\sigma$
Orionis brown dwarf with lithium in absorption, X-ray flares, and low-gravity
features (Burningham et al. 2005a; Caballero et al. 2006b).
We keep both S\,Ori~51 and S\,Ori~J053922.2--024552 in the sample. 

We have compared our selection criterion and the cluster sequence from our
data with available evolutionary tracks at very young ages. 
In particular, we have used the isochrones of the Lyon group (Baraffe et al.
1998 -- NextGen98; Chabrier et al. 2000 -- Dusty00; Baraffe et al. 2003 --
Cond03) at different ages and the most probable heliocentric distance of
$\sigma$~Orionis. 
The used magnitudes were those directly given by the models. 
The Dusty00 isochrone at 3\,Ma also acts as a good separation line at $I <$
20\,mag in the $I$ vs. $I-K_{\rm s}$ diagram. 
% CORYN: this sentence was written by the referee but does not sound very well
% for some of us...
Both with relation to the empirical cluster sequences and model isochrones
in Fig.~\ref{f_IvsIKs}, no blue outliers have been identified.

\section{Results and discussion}

\subsection{Individual masses}

Modelling ultracool atmospheres is a complicated issue, because it strongly
depends on the chemical composition, the dust content, the amount of
condensation and the size of the grains (Tsuji et al. 1996; Allard et al.
2001; Baraffe et al. 2002). 
To determine the masses of our objects, we will rely on the
mass-luminosity relations predicted by theoretical models of
their interior rather than on derived magnitudes in each passband.
Hence, we have derived the masses, M, of our $\sigma$~Orionis members and
candidates from their luminosities, $L$. 
In particular, we have used the M-$\log{\frac{L}{L_\odot}}$ relations of
the Lyon group and the basic properties of $\sigma$~Orionis provided in
Section~\ref{intro} (age, distance, metallicity). 
The mass-luminosity relations given by the NextGen98, Dusty00, and Cond03
models are identical.

The luminosity for each target was computed from the bolometric magnitude,
$M_{\rm bol} = M_{{\rm bol,}\odot} -2.5 \log{\frac{L}{L_\odot}}$ (where $M_{{\rm
bol,}\odot}$ = 4.74\,mag; Cox 2000).
The bolometric magnitude is: 

\begin{equation}
M_{\rm bol} = M_J^0 + BC_J 
\end{equation}

\noindent where $M_J^0 = M_J(J,d,A_J=0)$ is the absolute $J$-band magnitude
assuming no extinction and $BC_J$ is the bolometric correction in the $J$
band.
Caballero (2006) compiled from the literature $I$- and $J$-band photometry,
parallax measurements and luminosity determinations for all ultracool dwarfs
known at the time, from which he computed a bolometric correction in the $J$
band depending on the $I-J$ colour. 
The functional expression of $BC_J(I-J)$ was a cubic polynomial:

\begin{equation}
BC_J(I-J) = a_3 (I-J)^3 + a_2 (I-J)^2 + a_1 (I-J) + a_0
\end{equation}

\noindent with $a_3$ = +0.091\,mag$^{-2}$, $a_2$ = --0.875\,mag$^{-1}$, $a_1$ =
+2.486, $a_0$ = --0.140\,mag.
The standard deviation of the mean of the fit was $\sigma(BC_J)$ = 0.15\,mag.
The bolometric correction $BC_J$ is not strongly dependent on the $I-J$
colour.
It varies between 1.75$\pm$0.15 and 2.06$\pm$0.15\,mag in the 1.5\,mag $< I-J
<$ 4.0\,mag interval. 
The input data for this relation are the same as for other relations
between $BC_K$, $M_J$, $I-J$, $J-K$, $K-L'$ and spectral type of ultracool
dwarfs found in the literature (Dahn et al. 2002; Golimowski et al. 2004; Vrba
et al. 2004). 
Recent measurements of the bolometric magnitudes of ultracool dwarfs using
IRS/{\em Spitzer} spectra support previous $M_{\rm bol}$ determinations (Cushing
et al. 2006).
By using this $BC_J(I-J)$ relation, we assume that the spectral energy
distribution in field ultracool dwarfs and very young cluster members is
similar. 
The derivation of the $M_{\rm bol}$ from the $J$-band magnitude minimises
possible contributions to the total error by flux excesses at longer wavelengths
or photometric variability at bluer wavelengths (as in the case of T~Tauri stars
and substellar analogs).

The adopted distance modulus for the $\sigma$~Orionis cluster is $m-M$ =
7.78$\pm$0.42\,mag (Brown et al. 1994). 
The uncertainty of 20\,\% in the determination of the cluster distance ($d$
= 360$^{+70}_{-60}$\,pc), together with the uncertainty in the age 
(3$\pm$2\,Ma), are the most important contributors to the final error in the
mass of each target (the uncertainties in the distance may be even larger;
Caballero~2007). 
Errors coming from the photometric uncertainty or the use of the $BC_J(I-J)$
relation are comparatively smaller.
Theoretical models may also be a source of systematic uncertainty at very
young ages (Baraffe et al. 2002).
Although there is reasonable criticism on the validity of evolutionary tracks
at very low masses and very young ages (less than 10\,Ma), the observational
determination of the mass-age-luminosity triplet at different ages in the
substellar domain is in agreement with theoretical predictions (Bouy et al.
2004; Zapatero Osorio et al. 2004; Stassun, Mathieu \& Valenti 2006). 

We give in the last columns of Table~\ref{pre} the pair
M-$\log{\frac{L}{L_\odot}}$ for the 49 cluster members and member
candidates.  
The corresponding errors in mass and luminosity account for the propagation
of uncertainties in the $J$-band magnitude, and the age and distance of
$\sigma$~Orionis.
The theoretical effective temperatures derived from the
$\log{\frac{L}{L_\odot}}$-T$_{\rm eff}$ relation (between 3030$\pm$120 and
1740$\pm$70\,K) roughly match the expected effective temperatures of objects
with spectral type determination.  
Five objects have most probable masses larger than the hydrogen
burning mass limit and they are thus very low-mass stars.
Among the other 44 substellar objects, 11 are planetary-mass object
candidates and 33 are brown dwarf candidates.
The masses of some of our objects have been previously determined in
independent works (Zapatero Osorio et al. 2000; B\'ejar et al. 2001; Caballero
et al. 2006a; Gonz\'alez-Garc\'{\i}a et al. 2006), and differ by less than
10\,\% from our values.

\subsection{Infrared excesses and discs}
\label{section.discs}

\subsubsection{Detection of infrared excesses}

%______________________________________________ f_IR1IR4vsJKs
\begin{figure}
\centering
\includegraphics[width=0.49\textwidth]{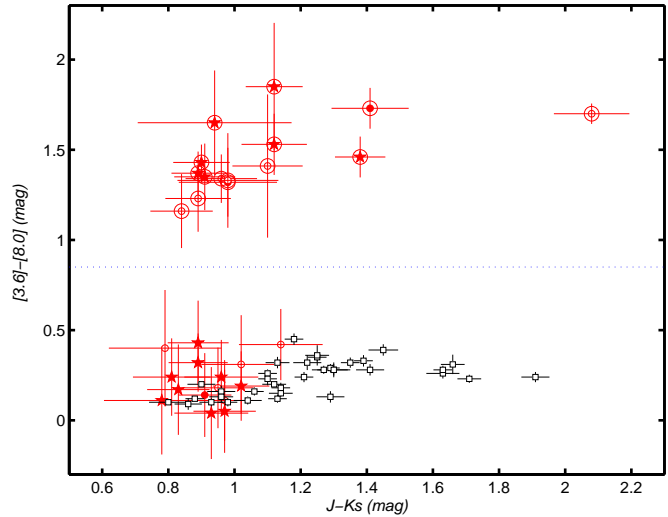}
\caption{$[3.6]-[8.0]$ vs. $J-K_{\rm s}$ colour-colour diagram. 
The symbols are as in Figs \ref{f_IvsIJ} and \ref{f_IvsIKs}.
The dotted line marks the boundary between objects with and without discs.
Field ultracool dwarfs with spectral types in the range M3.0--L5.0 from Patten
et al. (2006) are shown with open (black)~squares.}
\label{f_IR1IR4vsJKs}
\end{figure}
%

%______________________________________________ f_seds
\begin{figure*}
\centering
\includegraphics[width=0.99\textwidth]{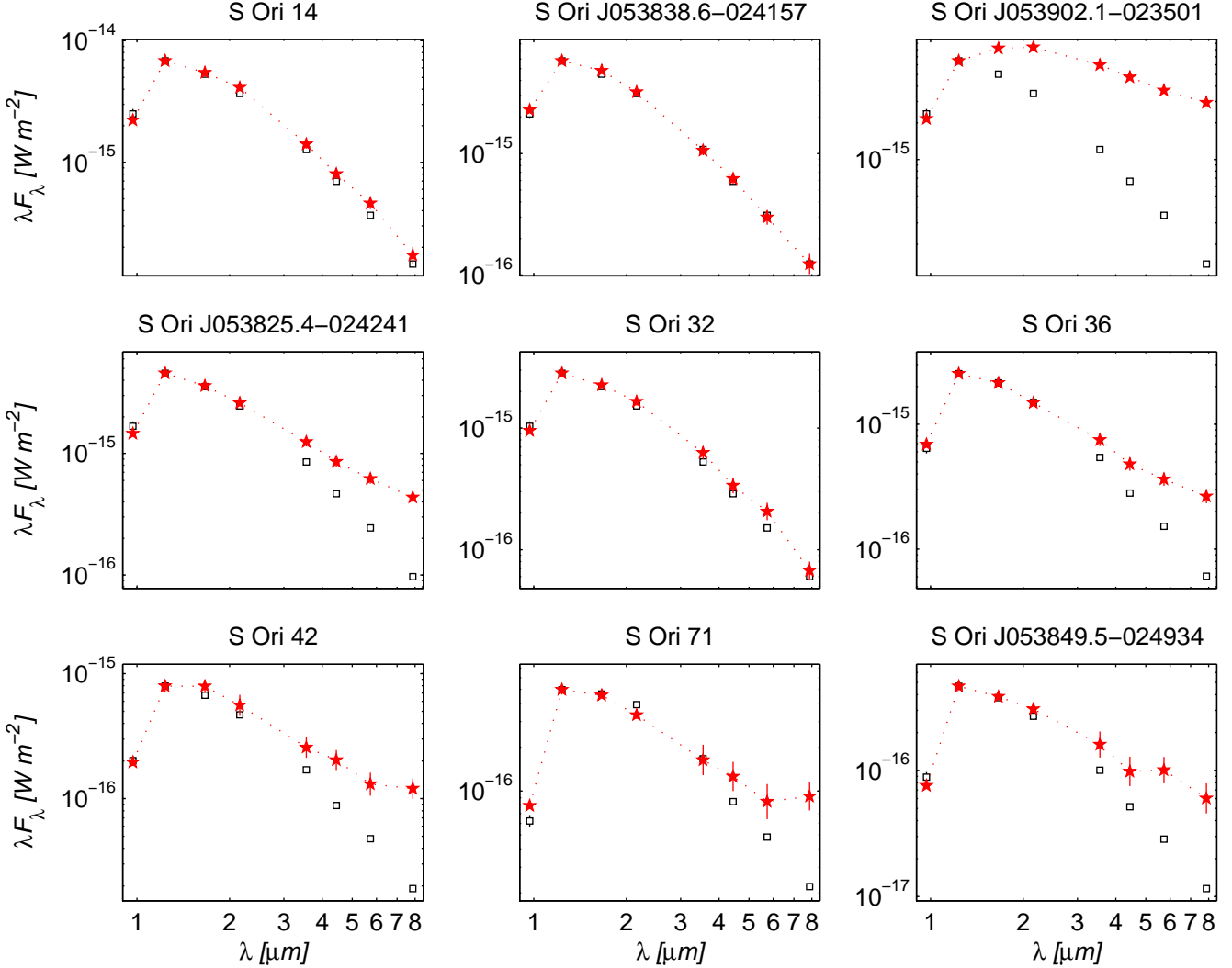}
\caption{Spectral energy distributions from the $I$ band to the $[8.0]$ band of
nine representative substellar objects in $\sigma$~Orionis (filled stars).
The open squares indicate the SEDs of ultracool field dwarfs of similar spectral
type, normalised at the $J$ band.}
\label{f_seds}
\end{figure*}
%
% It comes from \subsubsection{Properties of the disc harbours}

Brown dwarfs in very young regions have circum(sub)stellar discs in the
same way as T Tauri stars
(Wilking, Greene \& Meyer 1999; Natta \& Testi 2001; Fern\'andez \&
Comer\'on 2001; Jayawardhana, Mohanty \& Basri 2003; Furlan et al. 2005).  
With our follow-up, we have been able to measure flux excesses
redwards of 2\,$\mu$m in some of our targets, which are very probably
associated with the presence of circum(sub)stellar discs.
Objects with discs are extremely young ($\lesssim$ 10\,Ma), which confirms
membership in $\sigma$~Orionis\footnote{Jeffries et al. (2006) and
Caballero~(2007) have discussed the spatial overlapping of different young
populations of the Ori~OB~1b Association in the direction of $\sigma$~Orionis.}.
We have investigated different ISAAC/2MASS/IRAC colour-colour magnitude diagrams
to search for infrared excesses in our cluster members and candidates. 
The $[3.6]-[8.0]$ vs. $J-K_{\rm s}$ diagram (Fig.~\ref{f_IR1IR4vsJKs})
illustrates the difference in the colour between objects
without discs (0.0\,mag $\lesssim [3.6]-[8.0] \lesssim$ 0.5\,mag) and those that
very probably harbour discs ($[3.6]-[8.0] \gtrsim$ 1.0\,mag).  
The latter display flux excesses in the $[5.8]$ and $[8.0]$ bands with
respect to the $[3.6]$ band that are not found in any of the about 60
ultracool field dwarfs earlier than T3V studied with IRAC/{\em Spitzer} by
Patten et al.~(2006). 

Of the 30 cluster members and candidates with IRAC detection at
$[3.6]$ and $[8.0]$, 15 have colours $[3.6]-[8.0] >$ 0.90\,mag, which
we classify as objects with discs.  
Given the extremely red $K_{\rm s}-[8.0]$ colour of the confirmed cluster member
S\,Ori~J053847.2--025756, of 2.04$\pm$0.09\,mag, we will also classify it as an
object with disc, although it has only IRAC photometry at $[4.5]$ and
$[8.0]$. 
The 16 probable objects with discs are marked with a ``D'' in
Table~\ref{pre}.  
Our criterion is consistent with other disc selections in the literature
based on IRAC data.
In a very recent paper, Hern\'andez et al. (2007) study the spectral energy
distributions (SEDs) of most of our targets brighter than $J$ = 16.0\,mag. 
They classify the objects according to their $K_{\rm s}-[24]$  colours as brown dwarfs without excess, with optically-thick discs, and with
``evolved discs''.
We confirm all their disc detections for brown dwarfs brighter than $J$ =
16.0\,mag, and find a new one surrounding S\,Ori~J053847.2--025756, which is not
in their investigated area.
They also find a transition disc around S\,Ori~J053954.3--023719 (evident only
at the {\em Spitzer}/MIPS 24\,$\mu$m band) -- marked with ``(D)'' in
Table~\ref{pre}.

%__________________________________________________ One column table
   \begin{table*}
      \caption[]{Estimate of ultracool field dwarf contaminants.} 
         \label{number}
     $$ 
         \begin{tabular}{l c c c c c c c c c c c}
            \hline
            \hline
            \noalign{\smallskip}
$\Delta I$	& M3--5	& M5--7 & M7--9 & M9--L1 & L1--3 & L3--5 & L5--7 & L7--T0& T0--2 & T2--8 & All \\
(mag)   	&       &       &       &        &       &       &       &       &       &       & Sp. T. \\
            \noalign{\smallskip}
            \hline
            \noalign{\smallskip}
16.1--17.1	& 0	& 0.45  & 0.00  & 0      & 0     & 0     & 0     & 0     & 0     & 0     & 0.45 \\
17.2--18.2	& 0	& 0.25  & 0.03  & 0	 & 0	 & 0	 & 0	 & 0	 & 0	 & 0	 & 0.28 \\
18.3--19.3	& 0	& 0	& 0.14  & 0.01   & 0.00	 & 0	 & 0	 & 0	 & 0	 & 0	 & 0.15 \\
19.4--20.4	& 0	& 0	& 0.60  & 0.04   & 0.01	 & 0.00	 & 0	 & 0	 & 0	 & 0	 & 0.65 \\
20.5--21.6	& 0	& 0	& 0.74  & 0.17   & 0.05	 & 0.01	 & 0.00	 & 0	 & 0	 & 0	 & 0.97 \\
21.7--22.7	& 0	& 0	& 0.43  & 0.68   & 0.23	 & 0.05	 & 0.01	 & 0.00	 & 0	 & 0	 & 1.40 \\
22.8--23.8	& 0	& 0	& 0	& 0.86   & 0.78	 & 0.22	 & 0.06	 & 0.01	 & 0.01	 & 0	 & 1.94 \\
23.9--24.9$^{a}$& 0	& 0	& 0     & 0	 & 0.09	 & 0.28	 & 0.13	 & 0.02	 & 0.02	 & 0.01	 & 0.55 \\
            \noalign{\smallskip}
            \hline
            \noalign{\smallskip}
All intervals$^{a}$& 0	& 0.70	& 1.94	& 1.76   & 1.16	 & 0.56	 & 0.20	 & 0.03	 & 0.03  & 0.01	 & 6.39 \\
            \noalign{\smallskip}
            \hline
         \end{tabular}
     $$ 
\begin{list}{}{}
\item[$^{(a)}$] Assuming an incompleteness factor of 0.5 in the magnitude
interval 23.9--24.9\,mag. 
\end{list}
   \end{table*}
%
% It comes from \subsection{Contamination by fore- and background sources}

\subsubsection{Properties of the disc harbours}

In Table~\ref{spektra}, there are six cluster members that display strong and/or
broad H$\alpha$ emission. 
Three satisfy the accretion criterion of Barrado y Navascu\'es \& Mart\'{\i}n
(2003) (S\,Ori~J053825.4--024241, S\,Ori~42 and S\,Ori~71), while for 
the remaining three there are only qualitative estimates of the broadening of
the line. 
The width of the H$\alpha$ line at 10\,\% of the peak has been only measured for
S\,Ori~25 and S\,Ori~45, so the White \& Basri (2003) accretion criterion cannot
be applied here.
S\,Ori~71 is one of the strongest H$\alpha$ emitters close to the
deuterium-burning mass limit.
S\,Ori~J053825.4--024241 displays forbidden emission lines and is a very red
object ($J-K_{\rm s}$ = 1.38$\pm$0.08\,mag, $[3.6]-[8.0]$ =
1.46$\pm$0.11\,mag). 
There are hints for optical photometric variability in all six cases
(Caballero et al. 2004). 
S\,Ori~J053825.4--024241 and S\,Ori~42 are also variable in the $J$ band,
as discussed in Section~\ref{followup}.
All these objects, except S\,Ori~J054014.0--023127, which is not in the IRAC
survey area, have flux excesses at  $[5.8]$ and $[8.0]$ (and at $[24]$;
Hern\'andez et al. 2007).  
S\,Ori~J054014.0--023127 is a photometric variable and has an $I-J$ colour of
2.61$\pm$0.10\,mag, that is red for its $I$-band magnitude, and a marginally
broad H$\alpha$ emission (Kenyon et al. 2005). 
There are no H$\alpha$ measurements for the remaining cluster members
with~discs.

The SEDs of nine representative confirmed brown dwarfs, six with discs,
are shown in Fig.~\ref{f_seds}. 
The flux excesses at long wavelengths with respect to the expected fluxes from
ultracool dwarfs of roughly the same $I-J$ colours are evident in the six disc
harbours.
The comparison field dwarfs, whose infrared data have been taken from
Patten et al. (2006), are \object{GJ~1002} (M5.5), \object{DX~Cnc} (GJ~1111;
M6.5), V1054~Oph~E (\object{vB~8}, M7.0), and \object{2MASS
J12043036+3212595}~(M9.0).

\subsubsection{Frequency of brown dwarfs with discs}

According to the masses listed in Table~\ref{pre}, there are 2 stars
and 14 brown dwarfs with discs (as indicated by an excess emission at
8.0\,$\mu$m). 
Taking into account the completeness magnitude at $[8.0]$ and that objects
with discs have colours $[3.6]-[8.0] >$ 0.90\,mag, our 3.6--8.0\,$\mu$m
IRAC photometry is complete down to $\sim$0.015\,$M_\odot$, i.e. over most
of the brown dwarf mass interval.
We have derived the frequency of brown dwarfs with discs in $\sigma$~Orionis at 
47$\pm$15\,\% (14 brown dwarfs with infrared excess among 30 cluster members and
candidates with masses 0.072\,$M_\odot$ $>$\,M\,$>$ 0.015\,$M_\odot$ and
detection in the four IRAC channels). 
If we consider S\,Ori~J053954.3--023719 (with a transition disc according
to Hern\'andez et al. 2007) and some possible foreground ultracool contaminants,
the disc frequency could exceed~50\,\%.  

The value of 47$\pm$15\,\% is comparable to or slightly larger than 
other determinations of the frequency of discs surrounding stars and brown
dwarfs in the cluster. 
For example, the spectroscopically derived ratio of classical T~Tauri stars
to weak-line T~Tauri star in $\sigma$~Orionis is 30--40\,\% (Zapatero Osorio et
al. 2002a).
From infrared colours (observed flux excesses in the $L'$-band and/or in the
IRAC+MIPS/{\em Spitzer} passbands), Oliveira et al. (2006) and Hern\'andez et
al. (2007) found cluster disc frequencies of 27--39\,\% in the mass range
1.0--0.04\,$M_\odot$.  
Hern\'andez et al. (2007) did not observe any significant decrease in the disc
frequency towards the brown dwarf domain.  
Our result, which is among the first disc frequency determinations in the
mass interval 0.075--0.015\,$M_\odot$ (see the survey in Taurus by Luhman et al.
2006), combined with the cluster stellar data from the literature, suggests 
little dependence of the disc frequency with mass, from solar
masses down to 0.015\,$M_\odot$.
This supports theoretical scenarios where brown dwarfs form as a result
of an extension of the low-mass star formation process (see references in
Jameson~2005).

\subsection{Contamination by fore- and background sources}
\label{contamination}

Taking together the 20 confirmed cluster members in Table~\ref{spektra} and
the 18 cluster member candidates with discs in Table~\ref{pre}, from which eight
are also previously confirmed cluster members, then 30 objects of our 49
candidate cluster members exhibit signatures of extreme youth and, therefore,
are bona-fide cluster members. 
Out of the remaining 19 candidate cluster members without known youth features,
nine have low-resolution spectroscopy and ten have no spectroscopic
information or indications of discs from the IRAC photometry.
For two very faint objects there is no near-infrared follow-up at~all.

There could be contamination by red giants, galaxies and field dwarfs among
our targets awaiting membership confirmation.
All our cluster member candidates are far from the Galactic plane ($b$ =
--17.3\,deg) and have point-like PSF (i.e. they are non-extended), thus
contamination by red giants or galaxies is unlikely.
Besides, the 47 objects with near-infrared follow-up display colours that
match the dwarf sequence in a colour-magnitude diagram (e.g. $J-K_{\rm s}$ vs.
$I-J$ in Fig.~\ref{f_IvsIKs}).   
Therefore, we have estimated the back- and/or foreground contamination in our
$IJ$ survey only by non-member field-dwarf contaminants of very late
spectral types (intermediate- and late-M, L, and T).
Up-to-date models and data from the literature have been used:
($i$) a model of the Galactic thin disc by an exponential law (Phleps et al.
2005; Ryan et al. 2005; Karaali 2006);
($ii$) the length and height scales for late-type dwarfs in the Galaxy (Chen et
al. 2001); 
($iii$) the spatial densities, absolute magnitudes and colours of ultracool
dwarfs for each spectral type (Kirkpatrick et al. 1994; Dahn et al. 2002; Cruz
et al. 2003; Vrba et al. 2004; Nakajima~2005).

Table~\ref{number} shows the resulting possible contaminants listed in
approximately 1\,mag-wide bins. 
Since the last magnitude interval (23.9--24.9\,mag) is fainter than our
completeness limit, we have used an incompleteness factor (in particular,
only $\sim$50\,\% of the sources in this interval are detected
in the WFC data).  
The total number of possible field ultracool-dwarf contaminants in our survey is
$\sim$6.
Most of them are M7--L4V dwarfs in the magnitude interval $I$ =
20--24\,mag.  
The contribution to contamination by dwarfs later than L5 is very small. 
The figures reasonably match other determinations of the number of
contaminants along the line of sight to $\sigma$~Orionis (e.g. B\'ejar et al.
1999; Zapatero Osorio et al. 2002c; Gonz\'alez-Garc\'{\i}a et al. 2006).
Our contamination calculations predict $\sim$4 foreground ultracool dwarfs at
the magnitudes of the PMOs in the~cluster.

\subsection{The mass function in the substellar domain}

\begin{figure}
\centering
\includegraphics[width=0.49\textwidth]{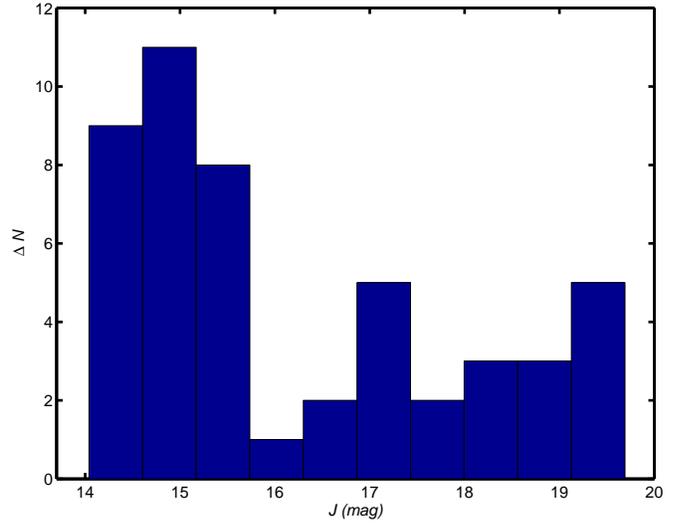}
\caption{Luminosity function in the $J$-band.
Note the scarcity of objects at $J \sim$ 16.0\,mag.}
\label{f_LFJ}
\end{figure}

\begin{figure}
\centering
\includegraphics[width=0.49\textwidth]{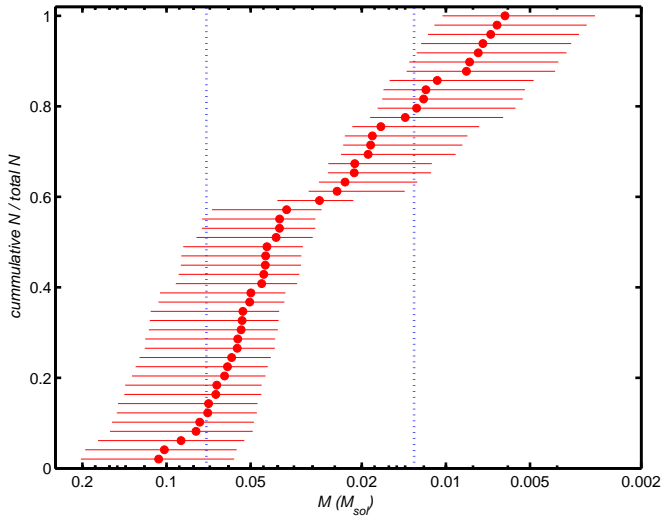}
\caption{Normalised cumulative number of objects ($\sum_i i/N$) vs. mass.
Filled circles indicate the most probable masses for 3\,Ma and 360\,pc.
The error bars indicate the masses for the extreme cases of a younger closer
cluster (1\,Ma and 300\,pc) and of an older farther cluster (5\,Ma and 430\,pc).
Dotted vertical lines denote the hydrogen (left) and deuterium (right) burning
mass limits.}
\label{f_NM}
\end{figure}

In Figs.~\ref{f_LFJ} and \ref{f_NM}, we show the luminosity function in $J$-band
and the normalised cumulative number of objects as a function of mass. 
There is an abrupt discontinuity in the distribution of magnitudes of brown
dwarfs at $J \sim$ 16.0\,mag, $I-J \sim$ 2.5\,mag, which translates into a
break in the mass distribution at $\sim$0.03\,$M_\odot$ for an age of 3\,Ma. 
In fact, there are only two cluster members with $J$-band magnitude
between 15.7 and 16.7\,mag. 

%______________________________________________ f_LFJ

\begin{figure}
\centering
\includegraphics[width=0.49\textwidth]{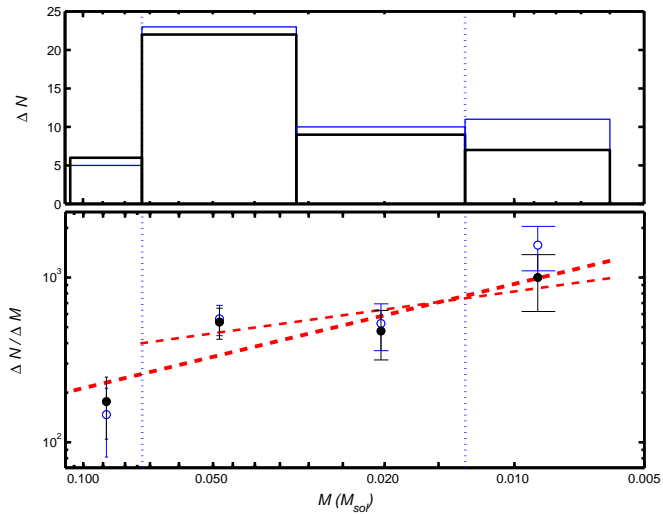}
\caption{Number of objects within each mass interval (top panel) and mass
spectrum (bottom panel) from 0.110 to 0.006\,$M_\odot$.
Dotted vertical lines denote the hydrogen (left) and deuterium (right) burning
mass limits.
The values  not corrected for contamination are shown with thin lines (top
panel) and open circles (bottom panel), while the values corrected for
contamination are shown with thick lines (top panel) and filled circles (bottom
panel). 
The fits of the mass spectrum in the three substellar bins (thin dashed
line) and the four bins (thick dashed line) are also shown.} 
\label{f_imf_sub}
\end{figure}

The number of objects within each mass interval and the mass spectrum ($\Delta N
/ \Delta {\rm M}$) from our data with and without contamination correction are
shown in Fig.~\ref{f_imf_sub}. 
The three mass intervals in the substellar regime are of the same width in
logarithmic scale, and classify our objects in high-mass brown dwarfs
(0.073--0.032\,$M_\odot$), low-mass brown dwarfs (0.032--0.013\,$M_\odot$) and
planetary-mass objects (0.013--0.006\,$M_\odot$).
The least-massive interval is complete down to 0.006\,$M_\odot$ for the
most probable age of the cluster.
These three mass intervals  are contaminated by 1, 1, and 4 field
ultracool dwarfs, respectively (see Section \ref{contamination}). 
On the contrary, the stellar mass domain (0.11--0.073\,$M_\odot$) has no
appreciable contamination. 
However, as mentioned in Section \ref{completeness}, the survey is not
complete at the stellar mass end due to saturation effects.
The survey area and the magnitude interval corresponding to very-low-mass
stars have been intensively investigated by other authors (B\'ejar et al.
1999, 2001; Scholz \& Eisl\"offel 2004; Sherry et al. 2004), and they did not
find any additional target fainter than $J$ = 14.2\,mag. 
We estimate that up to one cluster star between 0.11\,$M_\odot$ and the
hydrogen burning limit may have been missed.

%______________________________________________ f_NM

The best fit for the four de-contaminated intervals and for the most
probable age and distance gives an $\alpha$ power index of $\sim$0.65 
($\Delta N / \Delta {\rm M} \approx a {\rm M}^{-\alpha}$). 
Varying the widths of the mass bins, the $\alpha$ index changes
between 0.4 and 0.8.
We will assume $\alpha = +0.6\pm0.2$ as the slope of the mass spectrum in the
interval 0.11--0.006\,$M_\odot$. 
This value is similar to other determinations of the slope in the low-mass star
and substellar domain in $\sigma$~Orionis ($\alpha = +0.8\pm0.4$,  B\'ejar et
al. 2001; $\alpha = +0.6^{+0.5}_{-0.1}$, Gonz\'alez-Garc\'{\i}a et al. 2006). 
An extrapolation of the mass spectrum with index $\alpha = +0.6\pm0.2$ predicts
3--4 objects with masses 0.005--0.003\,$M_\odot$ in the area of our survey.
Taking into account only the three substellar mass intervals, the slope
decreases down to $\alpha = +0.4\pm0.2$.
It is equal within the error bars to the $\alpha$ value proposed by Kroupa
(2001) between 0.080 and 0.010\,$M_\odot$ ($\alpha = +0.3\pm0.7$).

%______________________________________________ f_imf_sub

Photometric variability, mass-segregation or unresolved binarity
corrections have not been applied in any mass interval:
\begin{itemize}
\item{Our $I$ and $J$ images were not taken simultaneously, so photometric
variability can potentially affect the selection procedure in the sense that
variable objects may be missed because they lie to the blue of our selection
criterion. 
Non-simultaneous photometric surveys in young regions in general, and in our
survey in particular, may yield incomplete lists of candidates.
However, as noted previously, our $J$-band images have been combined with
$I$-band data taken on two different occasions, and the resulting sample of
$\sigma$~Orionis candidates is not changed.
Furthermore, in Caballero et al. (2004) and in Scholz \& Eisl\"offel (2004) it
was found that the amplitude of the short- to long-term photometric variations
of $\sigma$~Orionis brown dwarfs is typically less than 0.15\,mag in the
$I$-band (only one source was observed with larger amplitude). 
Our selection criterion can account for such small variability, but does
not consider larger photometric variations, like the one possible observed in
S\,Ori J053948.1--022914 (see Section~\ref{022914}).
We estimate the incompleteness of our sample at about 2\,\% due to photometric
variability effects.
This rather small fraction does not seem to have a relevant impact on the study
of the cluster substellar mass funtion
(see Burningham et al. (2005b) for a discussion on apparent age spreads in OB
association colour-magnitude diagrams and variability).} 
\item{No significant mass-dependent difference between the spatial
distributions of cluster members below 5\,$M_\odot$ has been found in  
$\sigma$~Orionis (B\'ejar et al. 2004; Caballero, in prep.).
Only the most massive stars (M~$>$~5\,$M_\odot$) are preferentially found
towards the centre of the cluster.
In particular, the four most massive stars in $\sigma$~Orionis are in the
multiple stellar system that gives the name to the cluster (Caballero~2007).
Hillenbrand \& Hartmann (1998) found no evindence for mass segregation below
$\sim$1--2\,$M_\odot$ in the Orion Nebula Cluster (age $<$ 1\,Ma).
Given the extreme youth of the star-forming regions in Orion, there may not have
been enough time for mass segregation as in older well-known open clusters, as
the Pleiades or the Hyades (Pinfield et al. 1998; Perryman et al. 1998), or in
very old globular clusters.}
\item{Little is known about binarity in $\sigma$~Orionis, especially at
substellar masses.
Only a few works have found multiple systems in the cluster with low-mass
components (Lee et al. 1994 and Kenyon et al. 2005: spectroscopic binaries;
Caballero 2005 and Caballero et al. 2006b: wide binaries and companions).
Unresolved binaries of mass ratio close to 1 are $\sim$0.7\,mag brighter
than single objects, which leads to higher derived masses.
Some currently considered single low-mass stars could really be tight brown
dwarf pairs. 
However, more radial-velocity investigations and high spatial resolution imaging
are needed to quantify the binary frequency in $\sigma$~Orionis (see
K\"ohler et al. 2006 for a study of binary frequency in the Orion Nebula
Cluster).} 
\end{itemize}

Our list of cluster members may also be incomplete because the regions
around very bright stars within the field of view of the survey are not well
surveyed due to rather extensive glare that occupies several times the typical
FWHM of the $I$-band images.
However, the total lost area because of this effect is less than 2\,\% (see
previous Sections).
Given the spatial density of free-floating $\sigma$~Orionis members in the
investigated magnitude intervals (49 in 790\,arcmin$^2$, i.e.
0.062\,objects\,arcmin$^{-2}$), we estimate the number of missed cluster members
to be less than one.

Our mass spectrum goes to mass regimes not well covered in the literature.
Thus, it can only be compared with few mass spectra from previous works, like
those of Lucas \& Roche (2000), Najita et al. (2000), Muench et al. (2002), and
Slesnick, Hillenbrand \& Carpenter (2004). 
Given the low number of studied objects per mass bin (only a few times
over the Poissonian errors), the mass spectrum of the $\sigma$~Orionis
cluster is consistent with the mass spectra of other regions studied in the
literature.

The sharp drop observed at $J \sim$ 16\,mag (M $\sim$ 0.03\,$M_\odot$) in
the $\sigma$~Orionis luminosity function of Fig.~\ref{f_LFJ} is not explained by
any theoretical evolutionary model currently available in the literature
(Burrows et al. 1997; Baraffe et al. 1998, 2003; Chabrier et al. 2000).  
Similar drops or breaks at comparable magnitudes are also found in the 
near-infrared luminosity-functions of some recent deep searches conducted in
various young clusters and star-forming regions (Muench et al. 2003 in IC~348;
Pinfield et al. 2003 in the Pleiades and the Praesepe;
Lucas, Roche \& Tamura 2005 in the Orion Nebula Cluster; Bihain et al. 2006 in
the Pleiades).  
Muench et al. (2002) discussed an apparent break from a single power-law
decline of the Trapezium brown dwarf mass function around 0.02--0.03\,$M_\odot$
($K$ = 15.5\,mag). 
It stayed with a secondary peak of the mass function near the deuterium-burning
limit, followed by a rapid decline to lower masses.
Both the Trapezium and $\sigma$~Orionis clusters possess similar characteristics
(age, distance, metallicity, environment).
It is likely that Muench et al. (2002)'s break and ours share a common origin.
Dobbie et al. (2002) also found that the luminosity functions of star-forming
regions, open clusters and the field show a drop between spectral types M7
and M8, which corresponds to colours $I-J \sim$ 2.2--2.5\,mag.
As is apparent from Fig.~\ref{f_IvsIJ}, the sharp drop of the
$\sigma$~Orionis luminosity function also lies in this colour interval.
Dobbie et al. (2002) speculated that this is caused by the beginning of dust
formation in cool, ``neutral'' atmospheres.
To date, no satisfactory explanation exists to account for this feature. 

Finally, we have investigated the effects of the uncertainties in the cluster
age and distance on the mass spectrum. 
In particular, the slope of the mass spectrum is highly sensitive to age
variations, which prevents us from deriving the error in the $\alpha$ index with
an accuracy better than 0.2. 
The break at M $\sim$ 0.03\,$M_\odot$ did not affect our first $\alpha$
determination, since we used 0.032\,$M_\odot$ as the boundary between high- and
low-mass brown dwarfs in the mass spectrum. 
However, the break is displaced to the middle of the intervals in the mass
spectrum for the extreme cases of the cluster being younger and closer (1\,Ma
and 300\,pc) and older and farther (5\,Ma and 430\,pc).
For these cases, one of the two brown dwarf intervals is almost empty, and the
linear fit becomes meaningless.
Using more mass intervals does not solve the problem. 
A diagram with the normalised cumulative number of objects as a function
of mass, like the one shown in Fig.~\ref{f_NM}, is free of the
subjectiveness in the choice of mass intervals in the mass spectrum and appears
useful to compare the distribution of masses of stars and substellar objects in
different young clusters.
There is an apparent change of slope in the cumulative distribution of
Fig.~\ref{f_NM} at masses between 0.025 and 0.040\,$M_\odot$.
This is likely related to the sharp drop observed in the cluster luminosity
function. 

Our main goal was to investigate whether there is any feature in the mass
function that could be linked to the opacity mass-limit for objects forming via
fragmentation in molecular clouds.
At lower masses, our luminosity function is rather smooth with no evidence
for a theoretically predicted opacity mass-limit. 
We derived a most probable mass of 0.006\,$M_\odot$ for our least massive
cluster member candidate.
Besides, there are seven IPMO candidates with most probable masses in the
interval 0.008--0.006\,$M_\odot$.
This is a relatively large number, even accounting for possible contamination.
If there were an opacity limit just below the deuterium-burning mass limit, we
would expect a lower number of cluster member candidates in that mass interval,
which would lead to a steep drop in the last bin of the mass spectrum in
Fig.~\ref{f_imf_sub}.
Thus, if these objects form via fragmentation, any possible mass cut-off
of the mass function lies below 0.006\,$M_\odot$ (6\,M$_{\rm Jup}$). 
The smooth continuity of the substellar mass spectrum towards masses 
below the deuterium-burning limit may indicate that the formation of IPMOs
with M $>$ 0.006\,$M_\odot$ is merely an extension of that of brown dwarfs.
To determine the existence of the hypothetical opacity-limit cut-off, both
ultra-deep imaging with high sensitive facilities covering from 1 to 8\,$\mu$m
and intermediate-resolution spectroscopy are needed to detect fainter
isolated-planetary mass objects of a few Jupiter masses and assess their 
membership in the cluster.

\section{Summary}

We performed a 790\,arcmin$^2$-wide survey close to the centre of the
$\sigma$~Orionis cluster (3$\pm$2\,Ma; 360$^{+70}_{-60}$\,pc) in the $I$- and
$J$-bands, down to limiting magnitudes $I \sim$ 24.1\,mag and $J \sim$
21.8\,mag. 
We selected 49 objects from the $I$ vs. $I-J$ diagram.
Of them, 33 are brown dwarfs and 11 are planetary-mass objects at
the most probable age and distance of the cluster.
Twenty objects display spectroscopic features of youth.
Two brown dwarfs and two planetary-mass object candidates are reported here for
the first time.

The infrared follow-up from the $H$-band to 8.0\,$\mu$m with the 2MASS
catalogue, deep ground-based near-infrared imaging with 2- and 4-m-class
telescopes, and {\em Spitzer} Space Telescope archival infrared images did
not allow us to detect any blue interlopers in our sample.
However, it allowed us to identify 18 objects with flux excess at 
8.0\,$\mu$m that are likely to harbour discs.
Some of them are T~Tauri substellar analogs with discs.
The frequency of brown dwarfs with discs in $\sigma$~Orionis is estimated at
47$\pm$15\,\%.

Taking into account the spectroscopic information and the IRAC photometry, 
30 targets are confirmed as very young low-mass objects members of the cluster.
Among the remaining 19 cluster candidates, up to 6 could be foreground
contaminants (especially affecting the faintest magnitude interval in
our sample). 

We find a rising mass spectrum ($\Delta N / \Delta {\rm M} \propto {\rm
M}^{-\alpha}$) in the mass interval 0.11 and 0.006\,$M_\odot$.
The $\alpha$ index is +0.6$\pm$0.2 for the whole mass interval considered, and
+0.4$\pm$0.2 if we restrict it to the substellar domain between 0.073 and
0.006\,$M_\odot$.
A break in the luminosity function is apparent at about $J \sim$
16.0\,mag ($\sim$0.03\,$M_\odot$ at the most probable age of the cluster).  
Within the mass interval covered by our survey, there is no direct evidence
for the presence of an opacity-mass limit for objects formed via fragmentation
and collapse of molecular clouds.
Any possible mass cut-off would lie below 6\,M$_{\rm Jup}$ (0.006\,$M_\odot$).
Both brown dwarfs and IPMOs seem to form as an extension of the low-mass star
formation process.

\begin{acknowledgements}

We thank the anonymous referee for his/her detailed and careful report, J.
Hern\'andez for helpful discussion on {\em Spitzer} data, and A. Manchado, J.
A. Acosta and the rest of the LIRIS instrument team for acquiring some deep $H$
and $K_{\rm s}$ data. 
Partial financial support was provided by the Spanish Ministerio de
Ciencia y Tecnolog\'{\i}a proyect PNAYA 2006--12612 of the Plan Nacional
de Astronom\'{\i}a y Astrof\'{\i}sica. 
% ISAAC
Based on observations obtained at the Paranal Observatory, Chile, in ESO
program 68.C-0553(A).
% CFHTIR
Based on observations obtained at the Canada-France-Hawaii Telescope (CFHT)
which is operated by the National Research Council of Canada, the Institut
National des Sciences de l'Univers of the Centre National de la Recherche
Scientifique of France, and the University of Hawai'i. 
% Omega-2000
Based on observations collected at the Centro Astron\'omico Hispano Alem\'an (CAHA)
at Calar Alto, operated jointly by the Max-Planck Institut f\"ur Astronomie and
the Instituto de Astrof\'{\i}sica de Andaluc\'{\i}a (CSIC). 
% WFC2
Based on observations made with the Isaac Newton Telescope (INT) operated on the
island of La Palma by the Isaac Newton Group in the Spanish Observatorio del
Roque de Los Muchachos of the Instituto de Astrof\'{\i}sica de Canarias.
% CAIN-II
The Telescopio Carlos S\'anchez is operated on the island of Tenerife by the
Instituto de Astrof\'{\i}sica de Canarias in the Spanish Observatorio del Teide
of the Instituto de Astrof\'{\i}sica de Canarias.	
% Spitzer
This work is based in part on observations made with the {\em Spitzer} Space
Telescope, which is operated by the Jet Propulsion Laboratory, California
Institute of Technology under a contract with NASA.
% 2MASS
This publication makes use of data products from the Two Micron All Sky
Survey, which is a joint project of the University of Massachusetts and
the Infrared Processing and Analysis Center/California Institute of
Technology, funded by the National Aeronautics and Space
Administration and the National Science Foundation.
% IRAF
IRAF is distributed by National Optical Astronomy Observatories,
which are operated by the Association of Universities for Research in
Astronomy, Inc., under cooperative agreement with the National Science
Foundation.
% SIMBAD
This research has made use of the SIMBAD database, operated at CDS,
Strasbourg, France.  

\end{acknowledgements}

\appendix

\section{Other interesting sources}
\label{remarkable}

\subsection{S\,Ori~J053948.1--022914}
\label{022914}

S\,Ori~J053948.1--022914 was discovered by B\'ejar et al. (2001).
It is a M7.0$\pm$0.5 object with a stable chromospheric H$\alpha$ emission
(pEW(H$\alpha$) = --6.7$\pm$1.6\,\AA~; Barrado y Navascu\'es et al. 2003) and
with a very red $J-K_{\rm s}$ colour for its spectral type ($J-K_{\rm s}$ =
1.28$\pm$0.17\,mag). %  
It was proposed to be a X-ray source by Mokler \& Stelzer (2002) from {\em
ROSAT} data.
Flesch \& Hardcastle (2004) correlated, however, the X-ray source with a
brighter star at 10.2\,arcsec detected to the northeast in their APM/USNO-A
search.  
This star, \object{B 1.01--319}, displays low gravity features and radial
velocity consistent with membership in $\sigma$~Orionis (Burningham et al.
2005).
Therefore, the X-ray emission is likely ascribed to the star, at a projected
physical separation of about 3\,700\,AU to the M7.0 object.

Scholz \& Eisl\"offel (2004) detected a non-periodic photometric variability
with a {\em rms} of 0.139\,mag in the red-optical light curve of S\,Ori
J053948.1--022914.  
The average $I$ magnitude measured by them, 18.10\,mag, considerably differs
from other measurements in the literature ($I$ = 18.92\,mag in B\'ejar et al.
1999; $I$ = 18.67$\pm$0.07\,mag in B\'ejar et al. 2004; $I$ = 18.50\,mag in this
work). 
It might be a photometrically-variable cluster member.
Further studies are needed to assess its membership in $\sigma$~Orionis.

\subsection{New probable cluster non-members}
\label{newinteresting}

%__________________________________________________ Two column table
   \begin{table*}
      \caption[]{New probable cluster non-members.} 
         \label{interesting}
     $$ 
         \begin{tabular}{lccccccr}
            \hline
            \hline
            \noalign{\smallskip}
Id.	& $\alpha^{J2000}$ & $\delta^{J2000}$ & $I \pm \delta I$ (mag) &  $J \pm \delta J$ (mag) &  $H \pm \delta H$ (mag) &  $K_{\rm s} \pm \delta K_{\rm s}$ (mag) & Remarks	\\
            \noalign{\smallskip}
            \hline
            \noalign{\smallskip}
1	& 05 39 25.4 	& --02 29 43	& 23.11$\pm$0.18	& 20.08$\pm$0.10        & 19.43$\pm$0.07	& 19.07$\pm$0.07 & Early L-type field dwarf?	\\	
2	& 05 38 44.6 	& --02 44 40	& 23.8$\pm$0.5		& 20.64$\pm$0.08        & 19.90$\pm$0.09	& 19.22$\pm$0.10 & Late L-type field dwarf? 	\\	
3	& 05 39 59.3 	& --02 52 56	& 23.3$\pm$0.3		& 20.82$\pm$0.09        & 19.6$\pm$0.2		& 18.65$\pm$0.07 & Radio galaxy?		\\	
4	& 05 40 04.7 	& --02 29 27 	& $\gtrsim$24.5		& 21.56$\pm$0.10        & 20.01$\pm$0.12	& 18.52$\pm$0.13 & Extremely red galaxy? 	\\	
            \noalign{\smallskip}
            \hline
         \end{tabular}
     $$ 
   \end{table*}

In Table~\ref{interesting} we provide the coordinates and $IJHK_{\rm s}$
magnitudes of four interesting red objects that probably do not belong to the
$\sigma$~Orionis cluster.
They are bluewards of our selection criterion. 
Two of them, with identifications 1 and 2 in the Table, are likely L-type field
ultracool dwarfs in the foreground.
Their $J-K_{\rm s}$ and $I-J$ colours match the sequence of the ultracool field
dwarfs depicted in the colour-colour diagram in the bottom right panel in
Fig.~\ref{f_IvsIKs}.
The sources with identifications 3 and 4 are likely of extragalactic nature.
Interestingly, the object Id.~3 is located at only  1.5$\pm$1.0\,arcsec to the
radio source \object{TXS 0537--029} (Douglas et al. 1996).
Given the low spatial density of radio sources towards $\sigma$~Orionis, it is
possible that Id.~3 and the radio source are associated with a non-catalogued
elliptical galaxy 6.2\,arcsec to the southeast of Id.~3.

\section{On-line material}
\label{aphex}

% !!!
%
%{\it Note to the Editors: Tables \ref{pre} and \ref{nir} should go in
%the text body, not in this Appendix!
%However, they are too large and \LaTeX~shifts them after the Bibliography...
%Fig.~\ref{ds9} should only go in the electronic version.
%Appendix \ref{remarkable} could go as on-line material.}

%______________________________________________ ds9_IIR1IR4_CCD2N
\begin{figure*}
\centering
\caption{False-colour composite image of the northern part of survey area
corresponding to the detector CCD\#2 of the WFC (11\,arcmin wide; north is
up, east is left). 
Blue is for photographic $I$ (from the Canadian Astronomy Data Centre), green is
for IRAC 3.6\,$\mu$m and red is for IRAC 5.8\,$\mu$m.
The two very low-mass stars and the eleven brown dwarfs in the area are
marked with white circles.
The brightest blueish star is \object{HD~294278}, a K2-type foreground dwarf.
The remaining bright greenish and orange stars are known classical T Tauri
stars in $\sigma$~Orionis, like \object{TX~Ori} and \object{V505~Ori}.
NOT SHOWN IN THE ASTRO-PH VERSION. % !!!
}

\label{ds9}
\end{figure*}
%

%__________________________________________________ pre
   \begin{table*}
      \caption[]{Basic data of our $\sigma$ Orionis cluster members and candidates.} 
         \label{pre}
     $$ 
         \begin{tabular}{l c c c c c c c c}
            \hline
            \hline
            \noalign{\smallskip}
Name					& $\alpha^{J2000}$ & $\delta^{J2000}$ 	& $I$ (mag)	& $J$ (mag) 	& $\log{\frac{L}{L_\odot}}$ & Mass (M$_\odot$) & Sp.Type      & Remarks$^a$ \\  
            \noalign{\smallskip}
            \hline
            \noalign{\smallskip}
\object{S\,Ori J054000.2--025159}      & 05 40 00.15   & $-$02 51 59.4  & 16.19$\pm$0.04       & 14.04$\pm$0.08& --1.43$\pm$0.18       &   0.11$^{+0.10 }_{-0.05 }$ &		    & Y        \\ % 11142    Ca04
\object{S\,Ori J053848.1--024401}      & 05 38 48.19   & $-$02 44 00.8  & 16.15$\pm$0.03       & 14.09$\pm$0.08& --1.45$\pm$0.18       &   0.10$^{+0.09 }_{-0.05 }$ &		    & Y,D      \\ % 21184    S  
\object{S\,Ori J053833.9--024508}      & 05 38 33.88   & $-$02 45 07.8  & 16.15$\pm$0.03       & 14.24$\pm$0.07& --1.51$\pm$0.18       &   0.09$^{+0.09 }_{-0.04 }$ &		    & Y,D      \\ % 22241    S 
\object{S\,Ori J053911.4--023333}      & 05 39 11.40   & $-$02 33 32.8  & 16.35$\pm$0.04       & 14.38$\pm$0.11& --1.57$\pm$0.18       &  0.078$^{+0.08 }_{-0.03 }$ & M5.0$\pm$0.5  & Y        \\ % 33098    S
\object{S\,Ori 14}		       & 05 39 37.60   & $-$02 44 30.5  & 16.41$\pm$0.04       & 14.40$\pm$0.08& --1.58$\pm$0.18       &  0.076$^{+0.08 }_{-0.03 }$ &		    & Y        \\ % 42849    14  
\object{S\,Ori 11}		       & 05 39 44.33   & $-$02 33 02.8  & 16.29$\pm$0.04       & 14.50$\pm$0.12& --1.60$\pm$0.18       &  0.071$^{+0.08 }_{-0.02 }$ & M6.0$\pm$0.5  &	       \\ % 32653    11
\object{S\,Ori J054014.0--023127}      & 05 40 13.96   & $-$02 31 27.4  & 17.12$\pm$0.03       & 14.51$\pm$0.10& --1.60$\pm$0.18       &  0.071$^{+0.08 }_{-0.02 }$ &		    & Y        \\ % 32232 C, Ca04  
\object{S\,Ori J053847.2--025756}      & 05 38 47.15   & $-$02 57 55.7  & 16.74$\pm$0.02       & 14.54$\pm$0.08& --1.63$\pm$0.18       &  0.066$^{+0.07 }_{-0.02 }$ &		    & Y,D      \\ % 21798 C, Ca04 
\object{S\,Ori J053838.6--024157}      & 05 38 38.59   & $-$02 41 55.9  & 16.38$\pm$0.03       & 14.56$\pm$0.07& --1.63$\pm$0.18       &  0.067$^{+0.08 }_{-0.02 }$ & M5.5$\pm$1.0  & Y        \\ % 22097    S, Ca06
\object{S\,Ori 16}		       & 05 39 15.10   & $-$02 40 47.6  & 16.63$\pm$0.03       & 14.61$\pm$0.10& --1.66$\pm$0.18       &  0.062$^{+0.07 }_{-0.018}$ &		    &	       \\ % 41334 C, 16
\object{S\,Ori J053902.1--023501}      & 05 39 01.94   & $-$02 35 02.9  & 16.44$\pm$0.03       & 14.69$\pm$0.11& --1.67$\pm$0.18       &  0.061$^{+0.07 }_{-0.017}$ &		    & D        \\ % 33564    Ca04
\object{S\,Ori 25}		       & 05 39 08.95   & $-$02 39 58.0  & 16.95$\pm$0.03       & 14.70$\pm$0.10& --1.70$\pm$0.18       &  0.058$^{+0.07 }_{-0.016}$ & M6.5$\pm$0.5  & Y        \\ % 41080 C, 25 
\object{S\,Ori J053954.3--023719}      & 05 39 54.33   & $-$02 37 18.9  & 16.79$\pm$0.03       & 14.77$\pm$0.07& --1.72$\pm$0.18       &  0.056$^{+0.06 }_{-0.015}$ & M6.0$\pm$1.0  & (D)      \\ % 34451 C, C04 
\object{S\,Ori J053844.4--024037}      & 05 38 44.48   & $-$02 40 37.6  & 17.17$\pm$0.03       & 14.78$\pm$0.08& --1.72$\pm$0.18       &  0.056$^{+0.06 }_{-0.015}$ &		    & D        \\ % 21046    S, T 
\object{S\,Ori J053826.1--024041}      & 05 38 26.23   & $-$02 40 41.4  & 16.95$\pm$0.02       & 14.81$\pm$0.07& --1.74$\pm$0.18       &  0.054$^{+0.06 }_{-0.014}$ & M5.0$\pm$2.0  & Y        \\ % 23044 C, S 
\object{S\,Ori 20}		       & 05 39 07.58   & $-$02 29 05.6  & 16.80$\pm$0.03       & 14.83$\pm$0.10& --1.75$\pm$0.18       &  0.054$^{+0.06 }_{-0.014}$ & M5.5$\pm$0.5  &	       \\ % 31269 C, 20 
\object{S\,Ori J053829.0--024847}      & 05 38 28.97   & $-$02 48 47.3  & 16.81$\pm$0.02       & 14.84$\pm$0.09& --1.75$\pm$0.18       &  0.053$^{+0.06 }_{-0.014}$ & M6.0$\pm$0.5  & D        \\ % 23448 C, S 
\object{S\,Ori 27}		       & 05 38 17.42   & $-$02 40 24.3  & 17.05$\pm$0.03       & 14.92$\pm$0.08& --1.79$\pm$0.18       &  0.050$^{+0.06 }_{-0.013}$ & M6.5$\pm$0.5  & Y        \\ % 24047 C, 27 
\object{S\,Ori J053825.4--024241}      & 05 38 25.44   & $-$02 42 41.3  & 16.87$\pm$0.02       & 14.95$\pm$0.07& --1.79$\pm$0.18       &  0.050$^{+0.06 }_{-0.012}$ & M6.0$\pm$1.0  & Y,D      \\ % 23156 C, nIR 
\object{S\,Ori 28}		       & 05 39 23.19   & $-$02 46 55.8  & 17.14$\pm$0.03       & 15.12$\pm$0.16& --1.86$\pm$0.18       &  0.046$^{+0.05 }_{-0.012}$ &		    & Y        \\ % 43752 C, 28 
\object{S\,Ori 31}		       & 05 38 20.88   & $-$02 46 13.3  & 17.23$\pm$0.02       & 15.18$\pm$0.09& --1.89$\pm$0.18       &  0.044$^{+0.04 }_{-0.011}$ & M7.0$\pm$0.5  &	       \\ % 23341 C, 31 
\object{S\,Ori J053922.2--024552}      & 05 39 22.25   & $-$02 45 52.4  & 17.04$\pm$0.03       & 15.20$\pm$0.15& --1.89$\pm$0.18       &  0.044$^{+0.04 }_{-0.011}$ &		    &	       \\ % 43396 C, C04, GG06
\object{S\,Ori 30}		       & 05 39 13.08   & $-$02 37 50.9  & 17.29$\pm$0.03       & 15.20$\pm$0.07& --1.90$\pm$0.18       &  0.044$^{+0.04 }_{-0.011}$ & M6.0$\pm$0.5  & D        \\ % 34576 C, 30 
\object{S\,Ori 32}		       & 05 39 43.59   & $-$02 47 31.8  & 17.33$\pm$0.03       & 15.34$\pm$0.05& --1.95$\pm$0.18       &  0.041$^{+0.04 }_{-0.011}$ &		    & Y        \\ % 44072 C, 32 
\object{S\,Ori J053855.4--024121}      & 05 38 55.42   & $-$02 41 20.8  & 18.59$\pm$0.04       & 15.38$\pm$0.10& --1.88$\pm$0.18       &  0.045$^{+0.05 }_{-0.011}$ &		    & D,New    \\ % 41521 -- ?L?
\object{S\,Ori J054004.5--023642}      & 05 40 04.54   & $-$02 36 42.1  & 17.50$\pm$0.03       & 15.39$\pm$0.07& --1.97$\pm$0.18       &  0.039$^{+0.04 }_{-0.010}$ &		    & Y,D      \\ % 34252 C, S 
\object{S\,Ori 36}		       & 05 39 26.85   & $-$02 36 56.2  & 17.68$\pm$0.03       & 15.40$\pm$0.07& --1.97$\pm$0.18       &  0.039$^{+0.04 }_{-0.010}$ &		    & Y,D      \\ % 34332 C, 36 
\object{S\,Ori J053854.9--024034}      & 05 38 54.92   & $-$02 40 33.8  & 18.71$\pm$0.04       & 15.69$\pm$0.10& --2.02$\pm$0.18       &  0.037$^{+0.03 }_{-0.009}$ &		    & D,New    \\ % 41267 -- ?L? 
\object{S\,Ori J053918.1--025257}      & 05 39 18.13   & $-$02 52 56.3  & 18.64$\pm$0.04       & 16.12$\pm$0.08& --2.25$\pm$0.18       &  0.028$^{+0.012}_{-0.007}$ &		    & D        \\ % 11387 C, C04, GG06
\object{S\,Ori 42}		       & 05 39 23.41   & $-$02 40 57.6  & 19.05$\pm$0.03       & 16.49$\pm$0.10& --2.39$\pm$0.18       &  0.025$^{+0.006}_{-0.010}$ & M7.5$\pm$0.5  & Y,D      \\ % 41389 C, 42 
\object{S\,Ori 45}		       & 05 38 25.58   & $-$02 48 37.0  & 19.48$\pm$0.02       & 16.73$\pm$0.07& --2.47$\pm$0.18       &  0.023$^{+0.006}_{-0.010}$ & M8.5$\pm$0.5  & Y        \\ % 23438 C, 45 
\object{S\,Ori J053929.4--024636}      & 05 39 29.38   & $-$02 46 36.5  & 19.75$\pm$0.03       & 16.96$\pm$0.16& --2.56$\pm$0.18       &  0.021$^{+0.005}_{-0.010}$ &		    &	       \\ % 43639 C, C04, GG06
\object{S\,Ori 51}		       & 05 39 03.22   & $-$02 30 20.7  & 20.22$\pm$0.03       & 17.06$\pm$0.12& --2.55$\pm$0.18       &  0.021$^{+0.005}_{-0.010}$ & M9.0$\pm$0.5  & Y        \\ % 31624 C, 51 
\object{S\,Ori 71}		       & 05 39 00.30   & $-$02 37 06.7  & 20.03$\pm$0.03       & 17.23$\pm$0.07& --2.67$\pm$0.18       &  0.019$^{+0.005}_{-0.010}$ & L0.0$\pm$0.5  & Y,D      \\ % 34380 C, 71 
\object{S\,Ori J053849.5--024934}      & 05 38 49.59   & $-$02 49 33.3  & 20.08$\pm$0.02       & 17.30$\pm$0.08& --2.70$\pm$0.18       &  0.018$^{+0.005}_{-0.010}$ &		    & D        \\ % 21426 C, C04 
\object{S\,Ori 47}		       & 05 38 14.62   & $-$02 40 15.4  & 20.49$\pm$0.02       & 17.37$\pm$0.08& --2.68$\pm$0.18       &  0.019$^{+0.005}_{-0.010}$ & L1.5$\pm$0.5  &	       \\ % 24042 C, 47 
\object{S\,Ori 50}		       & 05 39 10.81   & $-$02 37 15.1  & 20.48$\pm$0.03       & 17.50$\pm$0.07& --2.75$\pm$0.18       &  0.017$^{+0.005}_{-0.009}$ & M9.0$\pm$0.5  &          \\ % 34433 C, 50 
\object{S\,Ori 53}		       & 05 38 25.12   & $-$02 48 02.7  & 20.93$\pm$0.02       & 17.92$\pm$0.06& --2.92$\pm$0.18       &  0.014$^{+0.005}_{-0.008}$ & M9.0$\pm$0.5  &	       \\ % 23417 C, 53 
\object{S\,Ori J053944.5--025959}      & 05 39 44.55   & $-$02 59 58.9  & 20.74$\pm$0.04       & 18.04$\pm$0.08& --3.00$\pm$0.18       &  0.013$^{+0.005}_{-0.007}$ &		    &	       \\ % 14218    C04roiz 
\object{S\,Ori J054007.0--023604}      & 05 40 06.95   & $-$02 36 04.7  & 21.14$\pm$0.03       & 18.22$\pm$0.07& --3.05$\pm$0.18       &  0.012$^{+0.005}_{-0.007}$ &		    &          \\ % 34081 D? C04roiz 
\object{S\,Ori J053956.8--025315}      & 05 39 56.83   & $-$02 53 14.4  & 21.25$\pm$0.04       & 18.27$\pm$0.08& --3.06$\pm$0.18       &  0.012$^{+0.005}_{-0.007}$ &		    &	       \\ % 11467 C, C04, GG06
\object{S\,Ori J053858.6--025228}      & 05 38 58.55   & $-$02 52 26.7  & 22.19$\pm$0.05       & 18.61$\pm$0.08& --3.12$\pm$0.18       &  0.011$^{+0.005}_{-0.006}$ &		    &          \\ % 11253 D? C, C04, GG06
\object{S\,Ori J053949.5--023130}      & 05 39 49.52   & $-$02 31 29.8  & 22.04$\pm$0.04       & 18.89$\pm$0.10& --3.29$\pm$0.18       &  0.008$^{+0.005}_{-0.004}$ &		    & New      \\ % 32240 C, sok 
\object{S\,Ori 60}		       & 05 39 37.50   & $-$02 30 41.9  & 22.52$\pm$0.05       & 19.05$\pm$0.10& --3.31$\pm$0.18       &  0.008$^{+0.005}_{-0.004}$ & L2.0$\pm$1.0  &	       \\ % 32003 C, 60 
\object{S\,Ori 62}		       & 05 39 42.05   & $-$02 30 32.3  & 22.52$\pm$0.05       & 19.18$\pm$0.10& --3.38$\pm$0.18       &  0.008$^{+0.005}_{-0.004}$ & L2.0$\pm$1.5  &	       \\ % 31669 C, 62 
\object{S\,Ori J053844.5--025512}      & 05 38 44.56   & $-$02 55 12.4  & 22.77$\pm$0.06       & 19.31$\pm$0.08& --3.42$\pm$0.18       &  0.007$^{+0.005}_{-0.004}$ &		    &	       \\ % 21672 C, nMR
\object{S\,Ori J054008.5--024551}      & 05 40 08.49   & $-$02 45 50.3  & 22.80$\pm$0.07       & 19.43$\pm$0.16& --3.48$\pm$0.18       &  0.007$^{+0.005}_{-0.004}$ &		    &	       \\ % 43377 -- *, nMR 
\object{S\,Ori J053932.4--025220}      & 05 39 32.42   & $-$02 52 20.3  & 22.83$\pm$0.07       & 19.54$\pm$0.08& --3.53$\pm$0.18       &  0.007$^{+0.004}_{-0.003}$ &		    &	       \\ % 11231 C, nMR 
\object{S\,Ori J054011.6--025135}      & 05 40 11.58   & $-$02 51 34.6  & 22.95$\pm$0.07       & 19.69$\pm$0.09& --3.59$\pm$0.18       &  0.006$^{+0.004}_{-0.003}$ &		    & New      \\ % 11047 -- ?? 
           \noalign{\smallskip}
            \hline
         \end{tabular}
     $$ 
\begin{list}{}{}
\item[$^{a}$] Entries for which there is a clear spectroscopic youth indicator
or evidence of disc, or are firstly identified in this work are marked by
``Y'', ``D'', and ``New'', respectively.  
\end{list}
   \end{table*}
%
%\object{S\,Ori J053948.1--022914}      & 05 39 48.26   & $-$02 29 14.5  & 18.50$\pm$0.03       & 16.47$\pm$0.09& --2.40$\pm$0.18       &  0.024$^{+0.006}_{-0.010}$ & M7.0$\pm$0.5  &	       \\ % 31301 C, S 
%
% En (Be04 y) tesis y no aqui:
% 	J053841.8-024455 (210229)
% 	J053836.2-024405 (220187)
%	J053845.0-025834 (210824)
% 	J053831.2-024525 (220264)

%__________________________________________________ pre
   \begin{table*}
      \caption[]{Near-infrared follow-up photometry of our $\sigma$ Orionis cluster members and candidates.} 
         \label{nir}
     $$ 
         \begin{tabular}{l c c c c c c c c c}
            \hline
            \hline
            \noalign{\smallskip}
Name			  & $J$ (mag)		  & $H$ (mag) 	  & $K_{\rm s}$ (mag) & $[3.6]$ (mag)	       & $[4.5]$ (mag)         & $[5.8]$ (mag)         & $[8.0]$ (mag)  & Flag$^{a}$	\\  
            \noalign{\smallskip}
            \hline
            \noalign{\smallskip}
S\,Ori J054000.2--025159  & 14.14$\pm$0.03	  & 13.47$\pm$0.02& 13.15$\pm$0.04    & 12.83$\pm$0.08         & 12.86$\pm$0.10        & 12.63$\pm$0.11        & 12.51$\pm$0.14 & 2224		\\ % 11142 Ca04, Ke05
S\,Ori J053848.1--024401  & 14.07$\pm$0.03	  & 13.58$\pm$0.03& 13.19$\pm$0.03    & 12.51$\pm$0.07         & 12.37$\pm$0.08        & 11.89$\pm$0.08        & 11.08$\pm$0.07 & 2224		\\ % 21184 Be04, Ke05
S\,Ori J053833.9--024508  & 14.25$\pm$0.03	  & 13.68$\pm$0.03& 13.35$\pm$0.04    & 12.81$\pm$0.08         & 12.64$\pm$0.09        & 12.33$\pm$0.10        & 11.44$\pm$0.09 & 2224		\\ % 22241 Be04, Ke05
S\,Ori J053911.4--023333  & 14.45$\pm$0.03	  & 13.93$\pm$0.03& 13.57$\pm$0.04    & 13.28$\pm$0.10         & 13.10$\pm$0.11        & 12.95$\pm$0.13        & 13.04$\pm$0.19 & 2224		\\ % 33098 Be01, ByN03, Ke05
S\,Ori 14		  & 14.38$\pm$0.03	  & 13.82$\pm$0.03& 13.38$\pm$0.03    & 13.02$\pm$0.09         & 12.92$\pm$0.10        & 12.78$\pm$0.12        & 12.83$\pm$0.17 & 2224		\\ % 42849 Be99, ByN03, Ke05
S\,Ori 11		  & 14.29$\pm$0.03	  & 13.72$\pm$0.03& 13.36$\pm$0.04    & 13.21$\pm$0.10         & 13.05$\pm$0.11        & 12.92$\pm$0.13        & 12.79$\pm$0.17 & 2224		\\ % 32653 Be99, Be01, ByN03 (Ha=-10.5+-1.5), Be04
S\,Ori J054014.0--023127  & 14.57$\pm$0.03	  & 13.98$\pm$0.03& 13.57$\pm$0.04    & --		       & --		       & --		       & --		& 2220		\\ % 32232 Ca04, Ke05
S\,Ori J053847.2--025756  & 14.52$\pm$0.03	  & 13.94$\pm$0.04& 13.46$\pm$0.04    & --		       & 12.62$\pm$0.09        & --		       & 11.42$\pm$0.08 & 2222		\\ % 21798 Ca04, Ke05
S\,Ori J053838.6--024157  & 14.56$\pm$0.03	  & 13.96$\pm$0.03& 13.65$\pm$0.04    & 13.33$\pm$0.10         & 13.20$\pm$0.10        & 13.25$\pm$0.15        & 13.19$\pm$0.21 & 2224		\\ % 22097 Be04, Ke05, Ca06
S\,Ori 16		  & 14.67$\pm$0.03	  & 14.04$\pm$0.04& 13.66$\pm$0.04    & 13.23$\pm$0.10         & 13.55$\pm$0.12        & 13.07$\pm$0.14        & 13.05$\pm$0.20 & 2224		\\ % 41334 Be99, Be01, ByN03, Ca04
S\,Ori J053902.1--023501  & 14.44$\pm$0.04	  & 13.38$\pm$0.03& 12.61$\pm$0.03    & 11.45$\pm$0.04         & 10.99$\pm$0.04        & 10.52$\pm$0.04        &  9.75$\pm$0.04 & 2224		\\ % 33564 Ca04
S\,Ori 25		  & 14.66$\pm$0.03	  & 14.14$\pm$0.04& 13.74$\pm$0.05    & 13.25$\pm$0.10         & 13.18$\pm$0.11        & 13.22$\pm$0.15        & 13.01$\pm$0.18 & 2224		\\ % 41080 Be99 (Ha), Be01, ByN03 (Ha), Ca04 (var), Mu03 (Ha, HeI), Ja03, Pa04, Ca04, Fr06 (X)
S\,Ori J053954.3--023719  & 14.75$\pm$0.03	  & 14.21$\pm$0.04& 13.80$\pm$0.05    & 13.36$\pm$0.10         & 13.30$\pm$0.12        & 13.36$\pm$0.16        & 13.25$\pm$0.20 & 2224		\\ % 34451 Ca04, Ca06 (Ha=-5+-1), [Be04]
S\,Ori J053844.4--024037  & 14.80$\pm$0.03	  & 14.21$\pm$0.04& 13.94$\pm$0.05    & 13.45$\pm$0.14         & 13.06$\pm$0.11        & 12.63$\pm$0.11        & 12.29$\pm$0.15 & 2224		\\ % 21046 Be04
S\,Ori J053826.1--024041  & 14.91$\pm$0.04	  & 14.28$\pm$0.04& 13.92$\pm$0.06    & 13.60$\pm$0.12         & 13.44$\pm$0.13        & 13.28$\pm$0.15        & 13.17$\pm$0.20 & 2224		\\ % 23044 ByN03, Ca04, Be04, Ke05, Ca06
S\,Ori 20		  & 14.96$\pm$0.04	  & 14.34$\pm$0.04& 13.90$\pm$0.05    & 13.52$\pm$0.11         & 13.49$\pm$0.13        & 13.24$\pm$0.15        & 13.48$\pm$0.23 & 2224		\\ % 31269 Be99, Be01, ByN03, SE04, Ke05
S\,Ori J053829.0--024847  & 14.82$\pm$0.04	  & 14.28$\pm$0.04& 13.88$\pm$0.06    & 13.16$\pm$0.09         & 12.85$\pm$0.10        & 12.46$\pm$0.10        & 11.82$\pm$0.10 & 2224		\\ % 23448 Be01, ByN03 (Ha-), Ca04
S\,Ori 27		  & 14.83$\pm$0.03	  & 14.31$\pm$0.04& 14.09$\pm$0.05    & 13.60$\pm$0.12         & 13.42$\pm$0.13        & 13.54$\pm$0.17        & 13.43$\pm$0.22 & 2224		\\ % 24047 Be99, Be01, ZO02, ByN03, Go03, Ca04, Be04, [ZO04], Pa05, Ke05, Pa05
S\,Ori J053825.4--024241  & 14.88$\pm$0.03	  & 14.16$\pm$0.04& 13.57$\pm$0.03    & 12.72$\pm$0.08         & 12.38$\pm$0.08        & 11.96$\pm$0.08        & 11.26$\pm$0.08 & 2224		\\ % 23156 Ca04, Be04, Ca06
S\,Ori 28		  & 15.33$\pm$0.04	  & 14.78$\pm$0.04& 14.34$\pm$0.07    & 13.91$\pm$0.13         & 13.87$\pm$0.16        & 13.80$\pm$0.20        & 13.80$\pm$0.27 & 2224		\\ % 43752 Be99, Be01, ZO02, ByN03, Ca04, [ZO04], Pa05, Ke05, Pa05
S\,Ori 31		  & 15.19$\pm$0.04	  & 14.57$\pm$0.05& 14.16$\pm$0.08    & 13.82$\pm$0.13         & 13.63$\pm$0.14        & 13.50$\pm$0.16        & 13.51$\pm$0.24 & 2224		\\ % 23341 Be99, BJ01 (var), Be01, ByN03 (Ha=-2.5+-0.9), [ZO03], [Jo03], Ca04, Be04, Pa04, Pa05
S\,Ori J053922.2--024552  & 15.32$\pm$0.04	  & 14.84$\pm$0.05& 14.41$\pm$0.08    & 14.23$\pm$0.16         & 13.99$\pm$0.16        & 13.94$\pm$0.21        & 13.83$\pm$0.28 & 2224		\\ % 43396 Ca04
S\,Ori 30		  & 15.24$\pm$0.04	  & 14.75$\pm$0.04& 14.31$\pm$0.07    & 13.77$\pm$0.12         & 13.59$\pm$0.14        & 13.22$\pm$0.15        & 12.54$\pm$0.14 & 2224		\\ % 34576 Be99, Be01, ByN03 (Ha=-16+-6), Ca04  	 
S\,Ori 32		  & 15.34$\pm$0.05	  & 14.78$\pm$0.06& 14.37$\pm$0.08    & 13.90$\pm$0.13         & 13.86$\pm$0.16        & 13.65$\pm$0.18        & 13.85$\pm$0.19 & 2224		\\ % 44072 Be99, Be01, ByN03, Ca04, Be04, Ke05
S\,Ori J053855.4--024121  & 15.62$\pm$0.10	  & 14.84$\pm$0.05& 13.97$\pm$0.06    & 12.88$\pm$0.08         & 12.52$\pm$0.08        & 12.08$\pm$0.09        & 11.15$\pm$0.08 & 2224		\\ % 41521 ---
S\,Ori J054004.5--023642  & 15.30$\pm$0.05	  & 14.81$\pm$0.05& 14.27$\pm$0.07    & 13.67$\pm$0.12         & 13.37$\pm$0.12        & 12.89$\pm$0.12        & 12.14$\pm$0.12 & 2224		\\ % 34252 Ca04, Be04, Ke05
S\,Ori 36		  & 15.46$\pm$0.04	  & 14.84$\pm$0.05& 14.49$\pm$0.06    & 13.71$\pm$0.12         & 13.48$\pm$0.13        & 13.04$\pm$0.13        & 12.36$\pm$0.14 & 2224		\\ % 34332 Be99, Be01, ByN03, Ca04, Ke05
S\,Ori J053854.9--024034  & 15.92$\pm$0.06	  & 15.17$\pm$0.06& 14.71$\pm$0.11    & 13.79$\pm$0.13         & 13.42$\pm$0.13        & 12.91$\pm$0.13        & 12.47$\pm$0.14 & 2224		\\ % 41267 ---
S\,Ori J053918.1--025257  & 16.15$\pm$0.08	  & 15.55$\pm$0.10& 15.14$\pm$0.13    & 14.47$\pm$0.17         & 14.20$\pm$0.18        & 13.86$\pm$0.20        & 13.14$\pm$0.20 & 2224		\\ % 11387 Ca04
S\,Ori 42		  & 16.72$\pm$0.13	  & 15.92$\pm$0.13& 15.55$\pm$0.21    & 14.87$\pm$0.21         & 14.41$\pm$0.20        & 14.15$\pm$0.23        & 13.22$\pm$0.20 & 2224		\\ % 41389 Be99, ByN03, Ca04
S\,Ori 45		  & 16.67$\pm$0.11	  & 16.02$\pm$0.13& 15.59$\pm$0.21    & 14.75$\pm$0.37         & 14.76$\pm$0.23        & 14.82$\pm$0.33        & $\gtrsim$14.8	& 2224		\\ % 23438 Be99, AMB00, BJ01, Be01, ZO02, ByN03, Mu03, ZO03, Ca04, [ZO04], [Pa05], [Pa05], [Mc05], [ByN05] -- H&Ks: 16.02$\pm$0.13& 15.59$\pm$0.21
S\,Ori J053929.4--024636  & 16.96$\pm$0.16	  & 16.52$\pm$0.11& 16.03$\pm$0.07    & 15.17$\pm$0.24         & 15.18$\pm$0.28        & 15.38$\pm$0.49        & $\gtrsim$14.8	& 1114		\\ % 43639 Ca04
S\,Ori 51		  & 17.06$\pm$0.12	  & 16.42$\pm$0.09& 16.13$\pm$0.10    & 15.44$\pm$0.27         & 15.31$\pm$0.30        & 15.16$\pm$0.41        & $\gtrsim$14.8	& 1114		\\ % 31624 ZO00, Be01, MS02, McG04, Ca04 
S\,Ori 71		  & 17.23$\pm$0.07	  & 16.52$\pm$0.11& 16.11$\pm$0.05    & 15.36$\pm$0.26         & 14.93$\pm$0.25        & 14.62$\pm$0.30        & 13.51$\pm$0.24 & 1114		\\ % 34380 ByN02, Liebert03, ByN03, Luhman03, ZO03, LM04, Ca04, ByN05, Lu05, Koen05, Ki05 
S\,Ori J053849.5--024934  & 17.30$\pm$0.08	  & 16.70$\pm$0.05& 16.20$\pm$0.07    & 15.38$\pm$0.26         & 15.20$\pm$0.29        & 14.43$\pm$0.26        & 13.97$\pm$0.30 & 1114		\\ % 21426 Ca04
S\,Ori 47		  & 17.53$\pm$0.24	  & 16.22$\pm$0.17& 15.81$\pm$0.08    & 14.94$\pm$0.21         & 14.96$\pm$0.26        & 14.75$\pm$0.30        & $\gtrsim$14.8	& 2234		\\ % 24042 Be99, ZO99, Basri00, ZO00, Ma01, Lu01, ByN01, ZO02, Foot02, ByN02, ZO02, ByN03, Jo03, McG04, Ca04, MH05, Koen05, Ki05
S\,Ori 50		  & 17.50$\pm$0.07	  & 17.12$\pm$0.14& 16.18$\pm$0.05    & 15.60$\pm$0.29         & 15.71$\pm$0.36        & 15.07$\pm$0.40        & $\gtrsim$14.8	& 1114		\\ % 34433 ZO00, Be01, ByN01, Ca04
S\,Ori 53		  & 17.92$\pm$0.06	  & 17.36$\pm$0.05& 16.76$\pm$0.07    & 15.94$\pm$0.34         & 15.96$\pm$0.41        & $\gtrsim$15.8	       & $\gtrsim$14.8	& 1114		\\ % 23417 ZO00, Be01, ByN01, Ca04
S\,Ori J053944.5--025959  & 18.04$\pm$0.08	  & 17.14$\pm$0.32& 16.98$\pm$0.12    & --		       & 16.61$\pm$0.20        & --		       & $\gtrsim$14.8	& 1102		\\ % 14218 Ca04, Beinprep
S\,Ori J054007.0--023604  & 18.22$\pm$0.07	  & 17.49$\pm$0.16& 16.88$\pm$0.12    & 16.07$\pm$0.20         & 16.10$\pm$0.20        & $\gtrsim$15.8	       & $\gtrsim$14.8	& 1104		\\ % 34081 Ca04, Beinprep
S\,Ori J053956.8--025315  & 18.27$\pm$0.08	  & 17.66$\pm$0.05& 17.18$\pm$0.07    & 16.55$\pm$0.20         & 16.47$\pm$0.20        & $\gtrsim$15.8	       & $\gtrsim$14.8	& 1114		\\ % 11467 Ca04
S\,Ori J053858.6--025228  & 18.61$\pm$0.08	  & 18.04$\pm$0.06& 17.29$\pm$0.07    & 16.29$\pm$0.20         & 15.96$\pm$0.20        & 15.88$\pm$0.60        & $\gtrsim$14.8	& 1114		\\ % 11253 Ca04
S\,Ori J053949.5--023130  & 18.89$\pm$0.10	  & 18.18$\pm$0.09& 17.42$\pm$0.20    & 16.57$\pm$0.30         & 16.62$\pm$0.20        & $\gtrsim$15.8	       & $\gtrsim$14.8	& 1114		\\ % 32240 C, sok
S\,Ori 60		  & 19.05$\pm$0.10	  & 18.00$\pm$0.50& 17.30$\pm$0.10    & 16.52$\pm$0.45         & 15.52$\pm$0.41        & $\gtrsim$15.8	       & $\gtrsim$14.8	& 1134		\\ % 32003 ZO00, Be01, Ma01, Lucas01, ByN01
S\,Ori 62		  & 19.18$\pm$0.10	  & 18.00$\pm$0.50& 17.86$\pm$0.10    & 16.99$\pm$0.34         & 16.83$\pm$0.48        & $\gtrsim$15.8	       & $\gtrsim$14.8	& 1134		\\ % 31669 ZO00, Be01, Ma01, ByN01
S\,Ori J053844.5--025512  & 19.31$\pm$0.08	  & 18.40$\pm$0.30& 17.99$\pm$0.07    & 16.82$\pm$0.51         & 16.79$\pm$0.54        & $\gtrsim$15.8	       & $\gtrsim$14.8	& 1114		\\ % 21672 Ca04, GG06
S\,Ori J054008.5--024551  & 19.43$\pm$0.16	  & --  	  & --  	      & --		       & --		       & --		       & --		& 1000		\\ % 43377 Ca04, GG06
S\,Ori J053932.4--025220  & 19.54$\pm$0.08	  & 18.70$\pm$0.06& 17.86$\pm$0.07    & 16.12$\pm$0.37         & 16.23$\pm$0.46        & $\gtrsim$15.8	       & $\gtrsim$14.8	& 1114		\\ % 11231 Ca04, GG06
S\,Ori J054011.6--025135  & 19.69$\pm$0.09	  & --  	  & --  	      & --		       & --		       & --		       & --		& 1000		\\ % 11047
\noalign{\smallskip}
            \hline
         \end{tabular}
     $$ 
\begin{list}{}{}
\item[$^{a}$] Source of the infrared photometry.
Three first digits ($JHK_{\rm s}$): 
0 -- not available;
1 -- this work;
2 -- 2MASS;
3 -- Mart\'{\i}n et al. 2001.
Fourth digit (IRAC passbands):
0 -- object in any IRAC channel area;
2 -- object in two IRAC channel areas;
4 -- object in four IRAC channel areas.
\end{list}
   \end{table*}
%
%S\,Ori J053948.1--022914  & 16.42$\pm$0.09	  & 15.59$\pm$0.10& 15.19$\pm$0.14    & 14.89$\pm$0.21         & 14.75$\pm$0.23        & 14.81$\pm$0.33        & $>$		& 2224		\\ % 31301 Be01, MS02 (X?), ByN03, SE04, Be04


\begin{thebibliography}{}

\bibitem[2003]{Alla01} Allard, F., Hauschildt, P. H., Alexander, D. R.,
Tamanai, A. \& Schweitzer, A. 2001, ApJ, 556, 357
\bibitem[1998]{Bara98} Baraffe, I., Chabrier, G., Allard, F., \& Hauschildt, P. H.
1998, A\&A, 337, 403 
\bibitem[2002]{Bara02} Baraffe, I., Chabrier, G., Allard, F., \& Hauschildt, P.
H. 2002, A\&A, 382, 563 
\bibitem[2003]{Bara03} Baraffe, I., Chabrier, G., Barman, T., Allard, F. \&
Hauschildt, P. H. 2003, A\&A, 402, 701 
\bibitem[2001]{Barr01b} Barrado y Navascu\'es, D., Zapatero Osorio, M. R.,
B\'ejar, V. J. S., Rebolo, R., Mart\'{\i}n, E. L., Mundt, R. \& Bailer-Jones, C.
A. L. 2001, A\&A, 377, L9 
\bibitem[]{Barr02a} Barrado y Navascu\'es D., Zapatero Osorio M. R., Mart\'{\i}n
E. L., B\'ejar V. J. S., Rebolo R. \& Mundt R. 2002, A\&A, 393, L85 
\bibitem[]{Barr03a} Barrado y Navascu\'es, D. \& Mart\'{\i}n, E. L. 2003, AJ,
126, 2997  
\bibitem[]{Barr03b} Barrado y Navascu\'es, D., B\'ejar, V. J. S., Mundt, R.,
Mart\'{\i}n, E. L., Rebolo, R., Zapatero Osorio, M. R. \& Bailer-Jones, C. A. L.
2003, A\&A, 404, 171 
\bibitem[]{Barr04b} Barrado y Navascu\'es, D., Stauffer, J. R., Bouvier, J.,
Jayawardhana, R. \& Cuillandre, J.-C. 2004, ApJ, 610, 1064 
\bibitem[]{Bate02} Bate, M. R., Bonnell, I. A. \& Bromm, V. 2002, MNRAS, 332, L65
\bibitem[]{Bate03} Bate, M. R., Bonnell, I. A. \& Bromm, V. 2003, MNRAS, 339, 577
\bibitem[2006]{Beau06} Beaulieu, J.-P., Bennett, D. P., Fouqu\'e, P. et al. 2006,
Nature, 439, 437
\bibitem[1999]{Beja99} B\'ejar, V. J. S., Zapatero Osorio, M. R. \& Rebolo, R. 1999,
ApJ, 521, 671 
\bibitem[2001]{Beja01b} B\'ejar, V. J. S., Mart\'{\i}n, E. L., Zapatero Osorio, M. R.,
Rebolo, R., Barrado y Navascu\'es, D., Bailer-Jones, C. A. L., Mundt, R., Baraffe,
I., Chabrier, C. \& Allard, F. 2001, ApJ, 556, 830 
\bibitem[2004]{Beja04b} B\'ejar, V. J. S., Zapatero Osorio, M. R. \& Rebolo, R.
2004, AN, 325, 705
\bibitem[2004]{Biha06} Bihain, G., Rebolo, R., B\'ejar, V. J. S., Caballero, J. A.,
Bailer-Jones, C. A. L., Mundt, R., Acosta-Pulido, J. A. \& Manchado Torres, A. 2006,
A\&A, 458, 805
\bibitem[2000]{Boss00} Boss, A. P. 2000, ApJ, 536, L101
\bibitem[1998]{Bouv98} Bouvier, J., Stauffer, J. R., Mart\'{\i}n, E. L., Barrado y
Navascu\'es, D., Wallace, B. \& B\'ejar, V. J. S. 1998, A\&A, 336, 490
\bibitem[2004]{Bouy04} Bouy, H., Duch\^ene, G., K\"ohler, R., Brandner, W., Bouvier, J.,
Mart\'{\i}n, E. L., Ghez, A., Delfosse, X., Forveille, T., Allard, F. et al. A\&A, 423,
341
\bibitem[1994]{Brow94} Brown, A. G. A., de Geus, E. J. \& de Zeeuw, P. T. 1994, A\&A,
289, 101 
\bibitem[2005]{Burn05a} Burningham, B., Naylor, T., Littlefair, S. P. \&
Jeffries, R. D. 2005a, MNRAS, 356, 1583
\bibitem[2005]{Burn05b} Burningham, B., Naylor, T., Littlefair, S. P. \&
Jeffries, R. D. 2005b, MNRAS, 363, 1389
\bibitem[1997]{Burr97} Burrows, A., Marley, M., Hubbard, W. B., Lunine, J. I.,
Guillot, T., Saumon, D., Freedman, R., Sudarsky, D. \& Sharp, C. 1997, ApJ, 491, 856 
\bibitem[2004]{Caba04a} Caballero, J. A., B\'ejar, V. J. S., Rebolo, R. \&
Zapatero Osorio, M. R. 2004, A\&A, 424, 857
\bibitem[2005]{Ca05} Caballero, J. A. 2005, Astron. Nachr., 326, No. 10, 1007
\bibitem[2006]{Ca06a} Caballero, J. A. 2006a, PhD thesis, Universidad de La Laguna
\bibitem[2006]{Caba06a} Caballero, J. A., Mart\'{\i}n, E. L., Zapatero Osorio, M. R.,
B\'ejar, V. J. S., Rebolo, R., Pavlenko, Ya. \& Wainscoat, R. 2006a, A\&A, 445, 143
\bibitem[2006]{Caba06b} Caballero, J. A., Mart\'{\i}n, E. L., Dobbie, P. D. \&
Barrado y Navascu\'es, D. 2006b, A\&A, 460, 635
\bibitem[2006]{Ca07} Caballero, J. A. 2007, A\&A, 466, 917
\bibitem[2000]{Chab00a} Chabrier, G. \& Baraffe, I. 2000, ARA\&A, 38, 337 
\bibitem[2000]{Chab00b} Chabrier, G., Baraffe, I., Allard, F. \& Hauschildt, P.
2000, ApJ, 542, 464 
\bibitem[2004]{Chau04} Chauvin, G., Lagrange, A.-M., Dumas, C., Zuckerman, B.,
Mouillet, D., Song, I., Beuzit, J.-L. \& Lowrance, P. 2004, A\&A, 425, L29 
\bibitem[2005]{Chau05b} Chauvin, G., Lagrange, A.-M., Dumas, C., Zuckerman, B.,
Mouillet, D., Song, I., Beuzit, J.-L. \& Lowrance, P. 2005, A\&A, 438, L25
\bibitem[2000]{Char00} Charbonneau, D., Brown, T. M., Latham, D. W. \& Mayor, M.
2000, ApJ, 529, L45
\bibitem[2001]{Chen01} Chen, B., Stoughton, C., Smith, J. A. et al. 2001, ApJ, 553,
184 
\bibitem[2001]{Cox00} Cox A. N. 2000, Allen's astrophysical quantities, 4th ed.
Publisher: New York: AIP Press; Springer, 2000. Editedy by A. N. Cox. ISBN:
0387987460  
\bibitem[2003]{Cruz03} Cruz, K. L., Reid, I. N., Liebert, J., Kirkpatrick, J. D. \&
Lowrance, P. J. 2003, AJ, 126, 2421
\bibitem[]{Cutr03} Cutri, R. M., Skrutskie, M. F., van Dyk, S. et al. 2003, VizieR
On-line Data Catalog: II/246.  
Originally published in: University of Massachusetts and Infrared Processing and
Analysis Center (IPAC/California Institute of Technology)
\bibitem[]{Cush06} Cushing, M. C., Roellig, T. L., Marley, M. S. et al. 2006, ApJ,
648, 614
\bibitem[]{Dahn02} Dahn, C. C., Harris, H. C., Vrba, F. J. et al. 2002, AJ, 124,
1170 
\bibitem[2002]{Dobb02} Dobbie, P. D., Pinfield, D. J., Jameson, R. F. \&
Hodgkin, S. T. 2002, MNRAS, 335, L79
\bibitem[1996]{Doug96} Douglas, J. N., Bash, F. N., Bozyan, F. A., Torrence, G.
W. \& Wolfe, C. 1996, AJ, 111, 1945
\bibitem[]{FeCo01} Fern\'andez, M. \& Comer\'on, F. 2001, A\&A, 380, 264
\bibitem[]{Fles04} Flesch, E. \& Hardcastle, M. J. 2004, A\&A, 427, 387	
\bibitem[]{Fran06} Franciosini, E., Pallavicini, R. \& Sanz-Forcada, J. 2006, A\&A,
446, 501
\bibitem[]{Furl05} Furlan, E., Calvet, N., D'Alessio, P. et al. 2005, ApJ, 621,
L129	
\bibitem[]{Goli04a} Golimowski, D. A., Leggett, S. K., Marley, M. S. et al. 2004,
AJ, 127, 3516  
\bibitem[]{GoGa06} Gonzalez-Garc\'{\i}a, B. M., Zapatero Osorio, M. R., B\'ejar, V.
J. S., Bihain, G., Barrado y Navascu\'es, D., Caballero, J. A. \&
Morales-Calder\'on, M. 2006, A\&A, 460, 799
\bibitem[]{Grav03} Greaves, J. S., Holland, W. S. \& Pound, M. W. 2003, MNRAS, 346,
441
\bibitem[2007]{Hdez07} Hern\'andez, J., Hartmann, L., Megeath, T., Gutermuth,
R., Muzerolle, J., Calvet, N., Vivas, A. K., Brice\~no, C., Allen, L., Stauffer, J. et
al. 2007, ApJ, accepted, eprint arXiv:astro-ph/0701476 
\bibitem[1998]{Hill98} Hillenbrand, L. A. \& Hartmann, L. W. 1998, ApJ, 492,
540
\bibitem[2005]{Jame05} Jameson, R. F. 2005, AN, 326, 874
\bibitem[2003]{Jaya03b} Jayawardhana, R., Mohanty, S. \& Basri, G. 2003, ApJ, 592,
282 
\bibitem[2006]{Jeff06} Jeffries, R. D., Maxted, P. F. L., Oliveira, J. M. \&
Naylor, T. 2006, MNRAS, 317, L6
\bibitem[2006]{Kara06} Karaali, S. 2006, AN, 327, 97
\bibitem[2005]{Keny05} Kenyon, M. J., Jeffries, R. D., Naylor, T., Oliveira, J. M.
\& Maxted, P. F. L. 2005, MNRAS, 356, 89 
\bibitem[1994]{Kirk94} Kirkpatrick, J. D., McGraw, J. T., Hess, T. R., Liebert, J. \&
McCarthy, D. W. Jr. 1994, ApJS, 94, 749 
\bibitem[2006]{Koeh06} K\"ohler, R., Petr-Gotzens, M. G., McCaughrean, M. J.,
Bouvier, J., Duch\^ene, G., Quirrenbach, A. \& Zinnecker, H. 2006, A\&A, 458, 461
\bibitem[2001]{Krou01} Kroupa, P. 2001, MNRAS, 322, 231
\bibitem[1992]{Land92} Landolt, A. 1992, AJ, 104, 340
\bibitem[1968]{Lee68} Lee, T. A. 1968, ApJ, 152, 913
\bibitem[1994]{Lee94} Lee, C. W., Mart\'{\i}n, E. L. \& Mathieu, R. D. 1994,
AJ, 108, 1445
\bibitem[2000]{Luca00} Lucas, P. W. \& Roche, P. F. 2000, MNRAS, 314, 858
\bibitem[2005]{Luca05} Lucas, P. W., Roche, P. F. \& Tamura, M. 2005, MNRAS,
361, 211
\bibitem[2000]{Luhm00b} Luhman, K. L. 2000, ApJ, 544, 1044 % 12/2000
\bibitem[2006]{Lu06} Luhman, K. L., Whitney, B. A., Meade, M. R., Babler, B.
L., Indebetouw, R., Bracker, S. \& Churchwell, E. B. 2006, ApJ, 647, 1180
\bibitem[]{Mart01b} Mart\'{\i}n, E. L., Zapatero Osorio, M. R., Barrado y
Navascu\'es, D., B\'ejar, V. J. S. \& Rebolo, R. 2001, ApJ, 558, L117 
\bibitem[]{Mart03b} Mart\'{\i}n ,E. L. \& Zapatero Osorio, M. R. 2003, ApJ, 593,
L113 
\bibitem[]{McGo04} McGovern, M. R., Kirkpatrick, J. D., McLean, I. S.,
Burgasser, A. J., Prato, L. \& Lowrance, P. J. 2004, ApJ, 600, 1020
\bibitem[]{Mokl02} Mokler, F. \& Stelzer, B. 2002, A\&A, 391, 1025
\bibitem[]{Mayo95} Mayor, M. \& Queloz, D. 1995, Nature, 378, 355
\bibitem[]{Muen02} Muench, A. A., Lada, E. A., Lada, C. J. \& Alves, J. 2002,
ApJ, 573, 366
\bibitem[]{Muen03} Muench, A. A., Lada, E. A., Lada, C. J., Elston, R. J.,
Alves, J. F., Horrobin, M., Huard, T. H., Levine, J. L., Raines, S. N. \&
Rom\'an-Z\'u\~niga, C. 2003, AJ, 125, 2029
\bibitem[]{Muze03} Muzerolle, J., Hillenbrand, L., Calvet, N., Brice\~no, C. \&
Hartmann, L. 2003, ApJ, 592, 266 
\bibitem[]{Naji00} Najita, J. R., Tiede, G. P. \& Carr, J. S. 2000, ApJ,
541, 977 
\bibitem[]{Naka05} Nakajima, T. 2005, IAC/TNG Workshop on Ultralow-mass star
formation and evolution, La Palma, 2005 June 28 -- July 1. Eds. E. L.
Mart\'{\i}n \& A. Magazz\`u 
\bibitem[]{Natt01} Natta, A. \& Testi, L. 2001, A\&A, 376, L22
\bibitem[]{Neuh05a} Neuh\"auser, R., Guenther, E. W., Wuchterl, G., Mugrauer, M.,
Bedalov, A. \& Hauschildt, P. H. 2005a, A\&A, 435, L13
\bibitem[]{Oasa06} Oasa, Y., Tamura, M., Nakajima, Y., Itoh, Y., Maihara, T.,
Iwamuro, F., Motohara, K., Hayashi, S. S., Hayashi, M. \& Kaifu, N. 2006, AJ, 131,
1608
\bibitem[]{Oliv02} Oliveira, J. M., Jeffries, R. D., Kenyon, M. J.,
Thompson, S. A. \& Naylor, T. 2002, A\&A, 382, L22
\bibitem[]{Oliv06} Oliveira, J. M., Jeffries, R. D., van Loon, J. Th. \&
Rushton, M. T. 2006, MNRAS, 369, 272
\bibitem[]{Patt06} Patten, B. M., Stauffer, J. R., Burrows, A. et al. 2006, ApJ,
651, 502
\bibitem[]{Perr98} Perryman, M. A. C., Brown, A. G. A., Lebreton, Y., G\'omez,
A., Turon, C., de Strobel, G. C., Mermilliod, J. C., Robichon, N., Kovalevsky, J. \&
Crifo, F. 1998, A\&A, 331, 81
\bibitem[]{Phle05} Phleps, S., Drepper, S., Meisenheimer, K. \& Fuchs, B. 2005,
A\&A, 443, 929
\bibitem[]{Pinf98} Pinfield, D. J., Jameson, R. F. \& Hodgkin, S. T. 1998,
MNRAS, 299, 955
\bibitem[]{Pinf03} Pinfield, D. J., Dobbie, P. D., Jameson, R. F., Steele, I.
A., Jones, H. R. A. \& Katsiyannis, A. C. 2003, MNRAS, 342, 1241
\bibitem[]{Rees76} Rees, M. J. 1976, MNRAS, 176, 483
\bibitem[]{Reip01} Reipurth, Bo \& Clarke, C. 2001, AJ, 122, 432
\bibitem[]{Ry05} Ryan, Jr. R. E., Hathi, N. P., Cohen, S. H. \& Windhorst, R. A.
2005, ApJ, 631, L159
\bibitem[]{Scal86} Scalo, J. M. 1986, Fund. Cos. Phys., 11, 1
\bibitem[]{Scho04} Scholz, A. \& Eisl\"offel, J. 2004, A\&A, 419, 249
\bibitem[]{Sher04} Sherry, W. H., Walter, F. M. \& Wolk, S. J. 2004, AJ, 128, 2316
\bibitem[]{Silk77} Silk, J. 1977, ApJ, 214, 152
\bibitem[]{Sles04} Slesnick, C. L., Hillenbrand, L. A. \& Carpenter, J. M.
2004, ApJ, 610, 1045
\bibitem[]{Stas06} Stassun, K. G., Mathieu, R. D. \& Valenti, J. A. 2006, Nature,
440, 311
\bibitem[]{Tohl80} Tohline, J. E. 1980, ApJ, 239, 417
\bibitem[]{Tsuj96} Tsuji, T., Ohnaka, K., Aoki, W. \& Nakajima, T. 1996, A\&A,
308, L29
\bibitem[2004]{Vrba04} Vrba, F. J., Henden, A. A., Luginbuhl, C. B. et al. 2004,
AJ, 127, 2948 
\bibitem[]{WhBa00} White, R. J. \& Basri, G. 2003, ApJ, 582, 1109
\bibitem[]{Wilk99} Wilking, B.~A., Greene, T. P. \& Meyer, M. R. 1999, AJ, 117, 469
\bibitem[]{Zap97a} Zapatero Osorio, M. R., Rebolo, R. \& Mart\'{\i}n,
E. L. 1997, A\&A, 317, 164 
\bibitem[]{Zap99b} Zapatero Osorio, M. R., B\'ejar, V. J. S., Rebolo, R.,
Mart\'{\i}n, E. L. \& Basri, G. 1999, ApJ, 524, L115 
\bibitem[]{Zapa00} Zapatero Osorio, M. R., B\'ejar, V. J. S., Mart\'{\i}n, E. L.,
Rebolo, R., Barrado y Navascu\'es, D., Bailer-Jones, C. A. L. \& Mundt, R. 2000,
Science, 290, 103 
\bibitem[]{Zapa02a} Zapatero Osorio, M. R., B\'ejar, V. J. S., Pavlenko, Ya.,
Rebolo, R., Allende Prieto, C., Mart\'{\i}n, E. L. \& Garc\'{\i}a L\'opez, R. J.
2002a, A\&A, 384, 937 
\bibitem[]{Zapa02b} Zapatero Osorio, M. R., B\'ejar, V. J. S., Mart\'{\i}n, E. L.,
Barrado y Navascu\'es, D. \& Rebolo, R. 2002b, ApJ, 569, L99 
\bibitem[]{Zapa02c} Zapatero Osorio, M. R., B\'ejar, V. J. S., Mart\'{\i}n, E. L.,
Rebolo, R., Barrado y Navascu\'es, D., Mundt, R., Eisl\"offel, J. \& Caballero, J. A.
2002c, ApJ, 578, 536 
\bibitem[]{Zapa03b} Zapatero Osorio, M. R., Caballero, J. A., B\'ejar, V. J. S.,
\& Rebolo, R. 2003, A\&A, 408, 663 
\bibitem[]{Zapa04b} Zapatero Osorio, M. R., Lane, B. F., Pavlenko, Ya.,
Mart\'{\i}n, E. L., Britton, M. \& Kulkarni, S. R. 2004, ApJ, 615, 958

\end{thebibliography}
\end{document}